\documentclass{aa}  

\usepackage{graphicx}
\usepackage{amsmath}
\usepackage{amssymb}
\usepackage{comment}
\usepackage{hyperref}

\usepackage{txfonts}
\newcommand{\s}{Sgr~A*}
\begin{document} 
\defcitealias{Palumbo:2023}{PWCJ}

\defcitealias{Johnson:2018}{J18}

   \title{Orbit design for mitigating interstellar scattering effects in Earth-space VLBI observations of Sgr A*}


   \author{Aditya Tamar
          \inst{\ref{inst1}}\thanks{Corresponding author; adityatamar@gmail.com}
          \and
          Ben Hudson\inst{\ref{inst2},\ref{inst2a}}\and Daniel C.M. Palumbo\inst{\ref{inst3},\ref{inst3a}}
          }

\institute{Department of Physics, National Institute of Technology Surathkal, Karnataka-575025, India\label{inst1} \and Faculty of Aerospace Engineering, Delft University of Technology, 2629HS Delft, The Netherlands \label{inst2} \and KISPE Limited, Farnborough, United Kingdom \label{inst2a} \and Center for Astrophysics | Harvard \& Smithsonian, 60 Garden Street, Cambridge, MA 02138, USA \label{inst3} \and Black Hole Initiative at Harvard University, 20 Garden Street, Cambridge, MA 02138, USA \label{inst3a} 
             }

   \date{Date}

 
  \abstract
       {The black hole Sagittarius A* (\s{}) is a prime target for next-generation Earth-space very-long-baseline interferometry missions such as the Black Hole Explorer (BHEX), which aims to probe baselines of the order of 20 G$\lambda$. At these baselines, \s{} observations  will be affected by the diffractive scattering effects from the interstellar medium (ISM). Therefore, we study how different parameter choices for turbulence in the ISM affect BHEX's observational capabilities to probe strong lensing features of \s{}. By using a simple geometric model of concentric Gaussian rings for \s{}'s photon ring signal and observing at 320 GHz, we find that the BHEX-ALMA baseline has the required sensitivity to observe \s{} for a broad range of values of the power-law index of density fluctuations in the ISM and the inner scale of turbulence. For other baselines with moderate sensitivities, a strong need for observations at shorter scales of $\approx13.5$ G$\lambda$ is identified. For this purpose, an orbit migration scheme is proposed. It is modeled using both chemical propulsion (CP)-based Hohmann transfers and electric propulsion (EP)-based orbit raising with the result that a CP-based transfer can be performed in a matter of hours, but with a significantly higher fuel requirement as compared to EP, which however requires a transfer time of around 6 weeks. The consequences of these orbits for probing \s{}’s spacetime is studied by quantifying the spatial resolution, temporal resolution and the angular sampling of the photon ring signal in the Fourier coverage of each of these orbits. We show that higher orbits isolate spacetime features while sacrificing both, signal lost to scattering and temporal resolution, but gain greater access to the morphology of the photon ring. Thus we find that orbits between the low earth regime and the reference BHEX orbit can provide rich access to Sgr A*’s parameter space.} 



   \maketitle
%

\section{Introduction}
The observations by the Event Horizon Telescope (EHT) collaboration have opened a new window to probe the strong gravity regime near black holes. Through the technique of very-long-baseline interferometry (VLBI) \citep{Thompson}, a virtual Earth-sized telescope has produced the near-horizon images of the Messier 87* (hereafter M87*) \citep{EHT_M87_I,EHT_M87_IV,EHT:2024} and Sagittarius A* (hereafter \s) \citep{EHT_Sgr_I} black holes. These images have improved our understanding of the interplay between lensing and polarisation \citep{EHT_M87_VII,EHT_M87_VIII,Goddi:2021}, discriminating magnetic field morphology near black holes \citep{Yuan:2022} and serving as an arbiter of supermassive black hole mass measurements \citep{EHT_M87_VI}. It is expected that the recently reported results by the EHT of observing M87* at 345 GHz \citep{Raymond:2024} will continue to stimulate research in these directions. 

The EHT plans to continue observing M87* and \s{} with improvements in both, software and instrumentation design \citep{EHT:2024_mid}. Also, there are planned ground-based expansions  that will increase the number of sites and expand their frequency coverage \citep{Doeleman:2023}. However, such terrestrial observations are subject to fundamental limitations. In particular, the maximum possible baseline length is restricted by the Earth's diameter. Therefore, since angular resolution $\theta$ is related to the observing wavelength $\lambda$ and distance $d$ between stations by the relation $\theta\approx\lambda/d$, for a fixed wavelength, longer baselines can only be achieved by adding a space-based orbiter. Indeed, at 320 GHz, assuming Earth's diameter to be $d\approx12.756\times10{^6}$ m, the maximum length of terrestial baselines is $\approx13.5$ G$\lambda$. The subsequent improvement in angular resolution is expected to help probe the black hole's photon ring \citep{Johnson:2020}, a strong lensing feature of optically thin accretion that is largely governed by the properties of the black hole spacetime and thereby encodes its mass and spin. However, as shown by \citep{Shavelle:2024} and \citep[][hereafter \citetalias{Palumbo:2023}]{Palumbo:2023}, it is possible that ground based polarimetric observations can begin to detect the presence of the photon ring, if not measure its morphology.

It is therefore natural that there have been several recent investigations for developing the science case for next-generation black hole imaging missions with a space-based orbiter, building upon lessons from prior space VLBI missions \citep{Gurvits:2020}. On one hand, there have been concepts proposed of performing space-space VLBI \citep{Roelofs:2019,Trippe:2023,Hudson_2023,Shlentsova:2024} with two or more orbiters that can offer extremely high angular resolution by forming baselines that are several multiples of those permitted by the Earth's diameter and perform observations at much higher frequencies that are also uncorrupted by the effects of the Earth's atmosphere. On the other hand, there are proposals of performing Earth-space VLBI  with one \citep{Likhachev:2022,Johnson:2024} or more \citep{Palumbo:2019,Fish:2020} space orbiters that offer the prospect of having improved angular resolution using long baselines, whilst also having a dense sampling of the $(u,v)$ coverage through the ground array. Furthermore, several complementary studies have been performed that highlight the utility of orbit design and optimization for performing VLBI observations of M87* and \s{} with a space-based component \citep{Fromm:2021,Andrianov:2021,Likhachev:2022,Tamar:2024A}. 

For this paper, the reference specifications will be of the Black Hole Explorer (BHEX) mission \citep{Johnson:2024} which aims to place an orbiter in a circular, polar orbit at an altitude of 20,192 km to achieve Earth-space baselines of the order of 20 G$\lambda$. The observations at such long baselines at a frequency of 240-320 GHz in its primary receiver are crucial for one of its main science goals of observing the photon ring of M87* and \s{}. The science goals of the mission as well as details of BHEX's proposed payload and spacecraft systems is given in \citep{Johnson:2024}.

There are several astrophysical motivations for performing near-horizon science with \s{}. Indeed, it has the largest ``shadow" size ($\sim\,50\mu$as) among all of the black holes observed by the EHT \citep{EHT:SgrA_III}. Of particular relevance for BHEX is the fact that since \s{}'s mass is well constrained to $\sim4.1$ million ${\rm M}_\odot$ \citep{Ghez:2008,GRAVITY:2019},  probing its photon ring can provide a spacetime-driven measurement of the black hole's spin \citep{Gralla:2020A,Broderick:2022,Avendano:2023}. 
However, its observations are complicated by several factors. In particular, \s{} has a distinct, anisotropic diffractive scattering component that diminishes the signal on long baselines \citep[][hereafter \citetalias{Johnson:2018}]{Johnson:2018}. This is tricky for missions such as BHEX whose proposed orbit lies at these baseline lengths \citep{Johnson:2024}. We do note that the RadioAstron mission has already made space-VLBI observations of \s{} \citep{Johnson:2021} but crucially no detections were reported on the baselines formed \textit{with} the Spektr-R orbiter. Moreover, the maximum baselines lengths probed were $\approx250$ M$\lambda$ which is around two orders of magnitude shorter than the ones accessible to BHEX.

Thus, while going to such long baselines may be useful for observing the photon ring of M87* \citepalias{Palumbo:2023}, the \s{} signal might actually be dominant on relatively shorter baselines beyond which the signal becomes prohibitively difficult to capture.

An additional concern in studies of the photon ring is the need to capture many realizations of the plasma configuration around the black hole. \s's mass, approximately 1500 times smaller than that of M87*\citep{EHT:SgrA_I}, sets a shorter dynamical timescale around the black hole, of order tens of minutes; for example, the innermost stable circular orbit for \s{} if it has no angular momentum is just thirty minutes. Thus, any interferometric experiment on \s{} that can incoherently average over varying source structure permits shorter VLBI observations, as even single nights contain many realizations of the accretion disks. This property creates challenges for instantaneous imaging, but favors a mission architecture in which shorter orbits are used for observation in sequence as the orbit is raised. In particular, lower orbits (i.e at altitudes lesser than that of the BHEX mission) can drastically improve the temporal resolution of the array, while suffering less from diffractive scattering effects, trading off in both cases against a greater typical contribution from the narrow photon ring on long baselines. Sampling a broad range of orbits is useful for mitigating unknown risks of interstellar scattering, as the parameters which most strongly determine how much signal is lost on long baselines are not well-constrained by ground data.

This paper is organised as follows. Section \ref{sec:scatt_vlbi_sgra} introduces the effects of scattering by the interstellar medium (ISM) on \s{} observations and through various characterizations of turbulence in the ISM. Section \ref{section:diff_scatt_pr} provides details of diffractive scattering effects on a simple geometric model used to model \s{}'s photon ring signature. Section \ref{sec:obs_sgra_es_vlbi} quantifies the sensitivity of BHEX baselines with three ground stations, namely ALMA, SMA and SMT, to observe \s{} for various values of the ISM turbulence parameters. Section \ref{sec:orb_transfers} introduces the orbit migration scheme aimed at making observations at baselines shorter than those provided by the final BHEX orbit. Section \ref{sec:practical_VLBI} studies the impact of maintaining a real-time downlink constraint and propulsion choices. Section \ref{sec:abc} discusses the consequences of the paper's orbit scheme for probing \s's accretion morphology and Section \ref{sec:conclusions} presents the Conclusions along with potential avenues for future work. An extended treatment of the formulae used to compute the propulsion-centric results are given in the Appendix along with Earth station specifications and an accessible description of some key terms used in this paper.


\section{Scattering effects on VLBI observations of Sgr A*} \label{sec:scatt_vlbi_sgra}
It is well established that the observations of \s{} at radio frequencies are affected by scattering effects due to the ionized ISM (\cite{Davies:1976,Lo:1993,Gwinn:2014,Johnson_GW:2015,Johnson:2016,Johnson:2016A}; \citetalias{Johnson:2018}; \cite{Psaltis:2018,Issaoun:2019,Zhu:2019,Cho:2022}). In the strong scattering regime (which is of relevance to radio wave scattering in the ISM), the scattering effects can be separated into two classes, namely
diffractive and refractive effects, that arise from ``small'' and ``large" scale phase gradients of the scattering screen
respectively \citep{Goodman:1989,Johnson:2016A}.   The effects of refractive scattering on \s{} observations have been studied by various authors (\cite{Johnson:2016}, \citetalias{Johnson:2018}, \cite{Issaoun:2019}). In this paper, we shall focus on diffractive scattering, since for sub-millimeter wavelengths, on long baselines, diffractive scattering suppresses the interferometric visibility amplitude\citep{Zhu:2019,EHT_SgrA_VII:2024}. 

The diffractive scattering effects are approximated as an ensemble average that acts as a convolution between the unscattered image and the scattering kernel, resulting in a ``blurred image'' \citep{Johnson:2016}. In the Fourier domain, which is relevant for VLBI observations, the convolution corresponds to a multiplication of the Fourier transforms of the unscattered image and the scattering kernel. For a detailed discussion of the various imaging regimes related to interferometric observations of scattering sources, see \citet{Narayan:1989}.

The parameters characterising the physical processes of the ISM have a strong impact on \s{}'s observations at radio frequencies. This is particularly apparent for turbulence in the ISM wherein \s{}'s observations have been used to constrain models for ISM turbulence with varying power-law spectral indices as well as the associated inner and outer scales (\citetalias{Johnson:2018};\citep{Issaoun:2019}). While such studies have focused on using refractive scattering as the main constraining tool, we work in the diffractive scattering regime and aim to understand how existing uncertainties in parameter values characterising ISM's turbulence processes affect BHEX's observations of \s{} on its longest baselines. The underlying scattering model used is from \citet{Psaltis:2018}, with reference model parameters for \s{} taken from \citetalias{Johnson:2018}.

Now, among the features specifying the scattering model, we focus on the slice of parameter space relevant to turbulence in the ISM. In particular, we consider the power-law index $\alpha$ of the phase structure function of the scattering screen and the inner scale of turbulence $r_{in}$, with all other values being the same as Table 3 in \citetalias{Johnson:2018}. Here $\alpha$ is related to the power-law index $\beta$ of the density fluctuations in the ISM by the relation $\alpha=\beta-2$ and $r_{\rm in}$, along with the outer scale $r_{\rm out}$, represents the scales over which fluctuations in the electron density follow an unbroken power law. Since the constraints on $\alpha$ and $r_{\rm in}$ have to be considered jointly \citepalias{Johnson:2018}, we will consider variations in their values in pairs that reproduce the long-wavelength apparent size of \s{}. The three branches of these parameters that we consider are:
\begin{enumerate}
    \item \underline{Set J}: $\alpha$=1.38 and $r_{\rm in}=800$ km,
    \\
    \item \underline{Set K}: $\alpha$=1.67 and $r_{\rm in}=600$ km,
    \\
    \item \underline{Set S}: $\alpha$=1.99 and $r_{\rm in}=1000$ km.
\end{enumerate} 
The values in Set J are those that are recommended by \citetalias{Johnson:2018}. The recommended set is not a strict prescription since the authors acknowledge that their constraint on $\alpha$ is ``somewhat indirect'' and the inner-scale value is ``likely'' to be $800$ km. Moreover, they note that the $\alpha$ value being different from the Kolmogorov value is inconsistent with other studies of the local ISM, pulsar broadening and VLBI studies of heavily scattered sources.  Thus, in Set K the index $\alpha$ takes the Kolmogorov value of $5/3$ (or equivalently $\beta=11/3$) and $r_{\rm in}=600$ km represents the ``robust'' lower limit found in \citetalias{Johnson:2018}. Indeed, there have been several studies that indicate strong support for a Kolmogorov power-law behaviour \citep{Cordes:1985,Armstrong:1995,Chepurnov:2010,Xu:2020} but observations of nearby pulsars \citep{Cordes:1986,Gupta:1993,Bhat:2004,Smirnova:2014,Filoth:2024} and VLBI observations of quasar B 2005+403 behind the Cygnus region in our Galaxy \citep{Gabanyi:2006} have shown signs of non-Kolmogorov scaling. Lastly, the values in Set S are sample values motivated by radio observations of blazar and pulsar sources by \citep{Tuntsov:2013} and also from Active Galactic Nuclei surveys (including \s) studying  large-scale scattering properties of the ISM in our Galaxy \citep{Koryukova:2022}. We also note that the authors in \citetalias{Johnson:2018} could not rule out larger inner scale values and so the choice of $r_{\rm in}=1000$ km for this set isn't at odds with our existing knowledge of VLBI observations \s{} at radio frequencies. The observational and physical motivations for having $\alpha<2$ (or equivalently $\beta<4$) along with the consequences of having an $\alpha=2$ spectra have been extensively discussed in \citet{Armstrong:1995}. We note that there have been Very Long Baseline Array observations \citep{Pushkarev:2013} that indicate evidence for having $\beta>4$ (and therefore $\alpha>2$) but we postpone an extended analysis of this parameter space for future work. However, we do note that diffractive effects dominate when the spectrum has $\beta<4$ whereas large-scale turbulent eddies with $\beta>4$ are expected to have more refraction \citep{Cordes:1986A}. The model with $\beta=4$ has been developed in detail in \citet{Lambert:2000}. Another parameter prescription of $\alpha=0$ is given by \cite{Goldrecih:1995} but their model prescription was ruled out by long-baseline observations with ALMA \citep{Issaoun:2019}. Therefore, we do not take this model into consideration.

We note that the goal of this paper is \textit{not} to perform a detailed study of the correct parameters specifying the ISM and its turbulent phenomenon. In the context of Earth-space VLBI observations of \s{}, we want to understand the extent to which uncertainties in the ISM models impacts BHEX's ability to observe the photon ring signature of \s{} in its proposed orbit. In other words, whilst recognizing the prevailing uncertainties in the ISM's parameter space, we wish to investigate whether BHEX can observe \s{} \textit{at all.} Nevertheless, we have still verified that the the parameter choices in sets J, K and S reproduce \s{}'s scattering features at lower frequencies thereby ensuring that our framework is not incompatible with radio observations of \s{} across frequencies \citep{Issaoun:2019}.

\section{Diffractive scattering and photon ring observations} \label{section:diff_scatt_pr}
In order to study the effects of diffractive scattering on photon ring observations, we consider a simple geometric model for the expected image morphology for \s{}. For characterising the photon ring structure in observations, we follow the nomenclature from existing literature of using the index $n$ which counts the number of half-orbits made by the photon (in the $\theta$ direction of the Boyer-Lindquist co-ordinate system) around the black hole \citep{Gralla:2020,Johnson:2020}. Here, the $n=0$ image is weakly lensed and arises from the ``direct'' emission while the $n=1$ image arises due to strong lensing by the photon making one half orbit around the black hole. It is this latter feature that is referred to as the ``n=1 photon ring'' but since we are not considering higher order images \citep{Avendano:2023,Avendano:2023A}, we'd simply refer to its signature as the ``photon ring''. Next, both $n=0$ and $n=1$ signatures are modeled as having a ring of finite thickness which in the Fourier domain is obtained by the convolution of an infinitesimally thin ring with a circular Gaussian kernel (see Appendix G of \cite{EHT_M87_IV} for further details). We also assume they have equal radii. Thus, the specifiable parameters for this model are the fluxes $F_{0},F_{1}$ and the thickness values $\sigma_{0},\sigma_{1}$ of the $n=0$ and $n=1$ rings respectively. The values chosen for these are:
\begin{enumerate}
    \item The fluxes $F_{0}$ and $F_{1}$ are taken to be 3.18 Jy and 0.31 Jy respectively. These values arise from fixing \s{}'s total flux to be in accordance with the results of \citet{Bower:2015} and from GRMHD simulations that indicate the photon ring to contribute $\approx10$\% of the total flux \citep{Ricarte:2015,Rosales:2021}.
    \item Both $n=0$ and $n=1$ rings are assumed to have an equal radius of 50 $\mu as$. This is in range of the diameter measurements made by the EHT \citep{EHT_Sgr_I}. The assumption of equal radii is motivated by both, studies of thin rings in General Relativistic Magnetohydrodynamic (GRMHD) simulations \citep{Tiede:2022} and covariant models of the accretion flow \citep{Ozel:2022}.
    \item The ring thickness $\theta$, or equivalently the Full Width of Half Maximum (FWHM) of the aforementioned Gaussian kernel, is taken to be 15 $\mu as$ and 1.5 $\mu as$ for the $n=0$ and $n=1$ rings respectively, with the relative ratio of $\approx10$\% being once again in accordance with GRMHD results \citep{Ozel:2022}. Note that the FWHM is related to the Gaussian thickness $\sigma$ by the relation $\theta=2\sqrt{2\ln2}\sigma$.
\end{enumerate}

\begin{figure*}
\centering
\includegraphics[width=\textwidth]{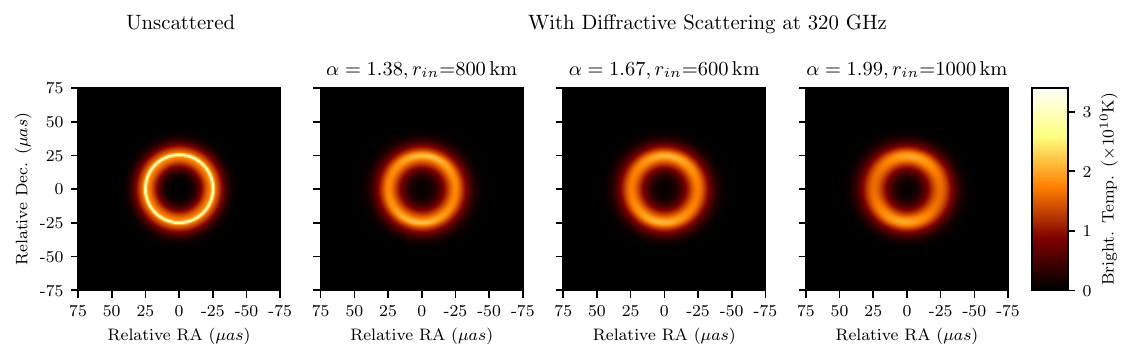}
\caption{The images of the geometric model of concentric Gaussian rings in the unscattered and diffractive scattering regime at 320 GHz. Here $\alpha$ is the power-law index of the phase structure function of the scattering screen and $r_{in}$ fixes the inner scale of the turbulence power-law regime. The parameter choices are discussed in Section \ref{sec:scatt_vlbi_sgra}.   }
\label{f:sgra_gmod_diff_scatt}
\end{figure*}

For these values and scattering parameters from Sets J, K and S from the previous Section, the unscattered and diffractive scattered image (under the ensemble-average regime) are shown in Figure \ref{f:sgra_gmod_diff_scatt}. The images are made using the \texttt{Stochastic Optics} package in \texttt{ehtim}.

\begin{figure}
\centering
\includegraphics[width=\columnwidth]{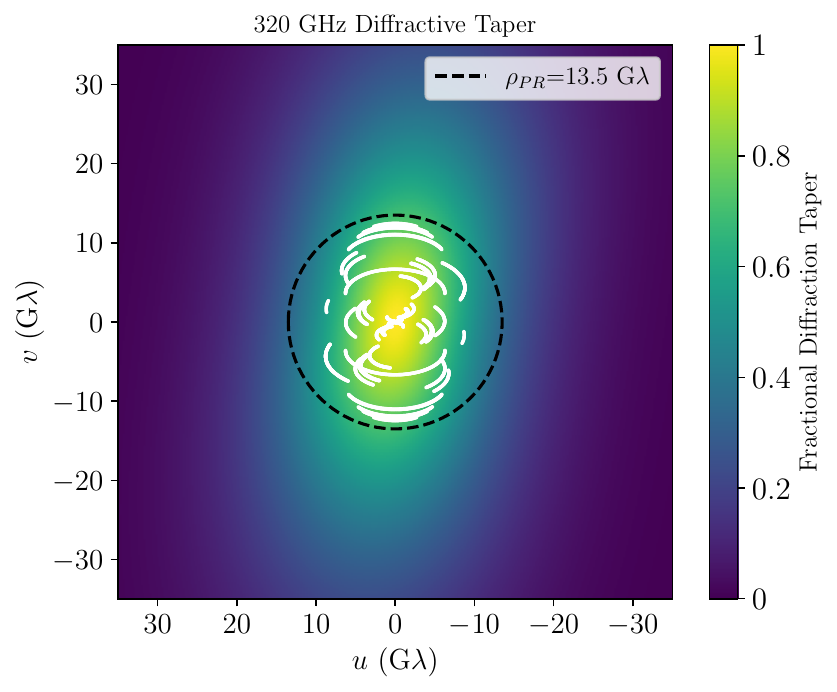}
\caption{The model of \s{}'s  diffractive scattering of \s at 320 GHz based on parameters from \citetalias{Johnson:2018}. Here $\rho_{PR}$ is the approximate $(u,v)$ radius predicted for our putative geometric model beyond which the photon ring signal is expected to dominate for \s's observations at the chosen frequency. The $(u,v)$ coverage in white is of the putative EHT 2025 array(see Table \ref{t:ground+dlink}).}
\label{f:Diff_Kernel}
\end{figure}

Now, it was recently shown by \cite{Tamar:2024} that for such a geometric model, closed-form expressions exist in total intensity, linear (LP) and circular polarisation that can specify the exact point in the Fourier domain where the photon ring signal first begins to dominate. Moreover, an analysis of the sensitivity and antenna diameter requirements by the authors indicated that for BHEX, accessing photon ring signatures in LP is much more likely than those from circular polarisation. The formula for the LP transition point, $(\rho_{T})_{LP}$, depends on the aforementioned parameters of the geometric model as well as on the ratio $\beta_{r}=\beta_{2,0}/\beta_{2,1}$ where the $\beta_{2}$ coefficient captures the rotationally symmetric polarisation structure \citep{Palumbo:2022}. The formula, reproduced from Equation 20 from \citet{Tamar:2024}, having $\mathcal{F}=F_{0}/F_{1}$ and $\sigma_{0}, \sigma_{1}$ being the Gaussian thickness for the $n=0$ and $n=1$ ring respectively, is given by,

\begin{gather} \label{eq:rhot_lp}
(\rho_{T})_{LP}\equiv\rho_{PR}=\sqrt{\frac{\ln(|\beta_{r}|\mathcal{F})}{2\pi^{2}(\sigma_{0}^{2}-\sigma_{1}^{2})}}.
\end{gather}
Substituting the aforementioned parameter values, along with a reasonable value of $\beta_{r}=3$ inferred from Figure 2 of \citep{Palumbo:2022}, Equation \ref{eq:rhot_lp} gives the transition point of $\approx 13.5\, G\lambda$ implying that under the assumptions of our geometric model, the photon ring signal starts to dominate after $13.5\, G\lambda$. This is consistent with the findings of \cite{Palumbo:2023} and would serve as an instructive reference value to gauge the baselines that need to be accessed by BHEX to probe \s's photon ring signal. This fact is also represented in Figure \ref{f:Diff_Kernel} wherein it is evident that the ground-based EHT array does not probe long enough baselines to be sensitive to the LP photon ring signal furnished by our model.

\section{Observing Sgr A* with Earth-space VLBI}\label{sec:obs_sgra_es_vlbi}
In this section, we quantify the sensitivity of BHEX baselines with the ground stations to probe the proposed geometric model of \s{}'s photon ring whilst considering variations in parameters of ISM turbulence across our three sets J ($\alpha=1.38$, $r_{\rm in}=800$ km), K ($\alpha=1.67$, $r_{\rm in}=600$ km) and S ($\alpha=1.99$, $r_{\rm in}=1000$ km).

\subsection{Instrumentation considerations: thermal noise}
The sensitivity of a baseline formed by two stations, say $i$ and $j$, is measured by the thermal noise $\sigma_{ij}$ for the baseline that is related to the stations' System Equivalent Flux Densities (SEFD), bandwidth $\Delta\nu$, integration time $t_{int}$ and quantisation efficiency $\eta$ by the relation,
\begin{gather}
    \sigma_{ij}=\frac{1}{\eta}\sqrt{\frac{\text{SEFD}_{i}\text{SEFD}_{j}}{2\Delta\nu t_{\text{int}}}}.
\end{gather}
The SEFD in turn depends on the system temperature $T_{\text{sys}}$, antenna efficiency $\eta_{A}$ and diameter $d$ as,
\begin{gather}
    \text{SEFD}=\frac{2k_{\text{B}}T_{\text{sys}}}{\eta_{A}\pi(d/2)^{2}},
\end{gather}
where $\pi(d/2)^{2}$ represents the antenna area and $k_{B}$ is the Boltzmann constant. For BHEX, assuming $T_{\text{sys}}=50$ K, $\eta_{A}=0.75$ and $d=3.5$ m \citep{Johnson:2024}, the SEFD value is, 
\begin{gather} \label{eq:SEFD_BX}
    \text{SEFD}\,(\text{BHEX})=17929.29\,\text{Jy}.
\end{gather}

Now for this paper, we consider BHEX's baselines with three stations: the Atacama Large Millimeter/submillimeter Array (ALMA) in Chile, the Submillimeter Array (SMA) facilities on Maunakea in Hawaii and the Submillimeter Telescope (SMT) in Arizona. These sites were part of the EHT array that in 2017 observed \s \citep{EHT_Sgr_I} and are also proposed to be part of the next generation EHT 
(ngEHT) with expanded multi-frequency capabilities \citep{Doeleman:2023}.

Our choice for the sites is governed by quantitative considerations. In particular, since $\sigma_{ij}$ is directly proportional to the SEFD at the station, a lower value of the site's SEFD would imply lesser thermal noise across the baseline thereby enhancing the sensitivity. Amongst our three sites, ALMA is the most sensitive, followed by SMA and then SMT. This is evident from the SEFD values,
\begin{gather}
    \text{SEFD}\,\text{(ALMA)}=74\,\text{Jy},\nonumber \\
    \text{SEFD}\,\text{(SMA)}=6700\,\text{Jy},\nonumber \\
    \text{SEFD}\,\text{(SMT)}=10500\,\text{Jy}, \label{eq:SEFD_Earth}
\end{gather}
taken from the specifications published by the EHT for their 2017 observing campaign \citep{EHT_M87_II}. As a consequence, the BHEX-ALMA baseline will be the most sensitive in the Earth-space array. We shall work with these SEFD values with the implicit assumption that any increase/decrease in SEFD values would decrease/increase the sensitivity of the corresponding baseline.

Now, using the SEFD values given in Equations \ref{eq:SEFD_BX} and \ref{eq:SEFD_Earth}, assuming BHEX  observes with a bandwidth of 16 GHz and integration time of 10 seconds, the thermal noise values $\sigma_{\text{BHEX-ALMA}},\sigma_{\text{BHEX-SMA}},\sigma_{\text{BHEX-SMT}}$ for our chosen baselines are,
\begin{gather}
    \sigma_{\text{BHEX-ALMA}}=2.71\, \text{mJy}, \nonumber\\
    \sigma_{\text{BHEX-SMA}}=25.83\, \text{mJy}, \nonumber\\
    \sigma_{\text{BHEX-SMT}}=32.33\, \text{mJy}.
\end{gather}
These $\sigma$ values would serve as the ``floor'' such that the amount of signal that can be probed by these baselines must have flux values higher than the detection threshold of $3\sigma$. Note that there is an unfortunate overlap of notation of using $\sigma$ for representing both, the Gaussian thickness of the rings and the thermal noise for a baseline. Nevertheless, one can refer to the subscripts to clarify the context of their usage; for the discussion on thermal noise floor, $\sigma$ represents the thermal noise values for the baselines considered above.

\subsection{Results}
For the BHEX-ALMA, BHEX-SMA and BHEX-SMT baselines, observations of our model of \s's signal are shown in Figure \ref{f:sgra_bsl_alphas}. The gray region represents the inaccessible flux that lies below the thermal noise floor for the corresponding baseline. Conversely, the colored region represents the flux that lies above the floor. The red dots represent the $(u,v)$ coverage for the corresponding baseline for 24 hours of observation, sampled at intervals of 60 minutes.

\subsubsection{Influence of the scattering parameters }
The dependence of the accessible signal on the scattering parameters $\alpha$ and $r_{\rm in}$ are quite evident. For any baseline, horizontally traversing towards higher values of $\alpha$ leads to lesser amount of signal above the thermal noise floor, consequently limiting the ability to observe \s on the longest baselines. We've also checked that the effect of varying $r_{\rm in}$ within our range is not as severe as that from the variations in $\alpha$. More broadly, this implies that the existing uncertainty over the properties of the ISM turbulence is intimately related to BHEX's ability to observe \s and therefore requires careful consideration. Furthermore, a column-wise comparison makes it evident that if the model parameters presented by \citetalias{Johnson:2018} are indeed accurate, a greater quanta of signal is accessible on the longest baselines.

\subsubsection{Baseline considerations}
From the first row of the Figure, it is encouraging to see that even if we move towards the limiting value of $\alpha_{S}=1.99$, the sensitivity of the BHEX-ALMA baseline will continue to probe the region where \s{}'s signal persists. Therefore, at least on its most sensitive baseline, for the parameter space being considered in this paper, BHEX should observe \s. 

For the BHEX-SMA and BHEX-SMT baselines, the situation is a little more challenging. Firstly, higher values of $\alpha$ lead to the source flux becoming less accessible to long baseline observations, with the signal sparsity being particularly acute in the BHEX-SMT baseline. However, note that in both of these cases (and of course for the BHEX-ALMA baseline as well), the region until $\approx13.5\, G\lambda$ continues to have accessible flux. Therefore, we make the inference that for BHEX to work with an array of Earth stations with varying levels of baseline sensitivities, the suitable region of the visibility domain for \s{} observations actually lies around $13.5\,G\lambda$. This is shorter than the currently envisaged $20\, G\lambda$ baseline lengths envisioned by the mission (when observing at 320 GHz)\citep{Johnson:2024}.

Keeping these considerations in mind, in the subsequent sections we shall lay out a dynamic orbit scheme, quantified by mission design considerations, that can potentially allow BHEX to still go to its proposed final orbit but also pass through and perform observations at intermediate orbits. Such an implementation would provide access regions of the Fourier space identified above which would continue to have accessible signal for baselines of varying sensitivities, while mitigating the potential loss in signal due to uncertainties in the ISM's turbulence properties.
\begin{figure*}
\centering
\includegraphics[width=\textwidth,height=14cm]{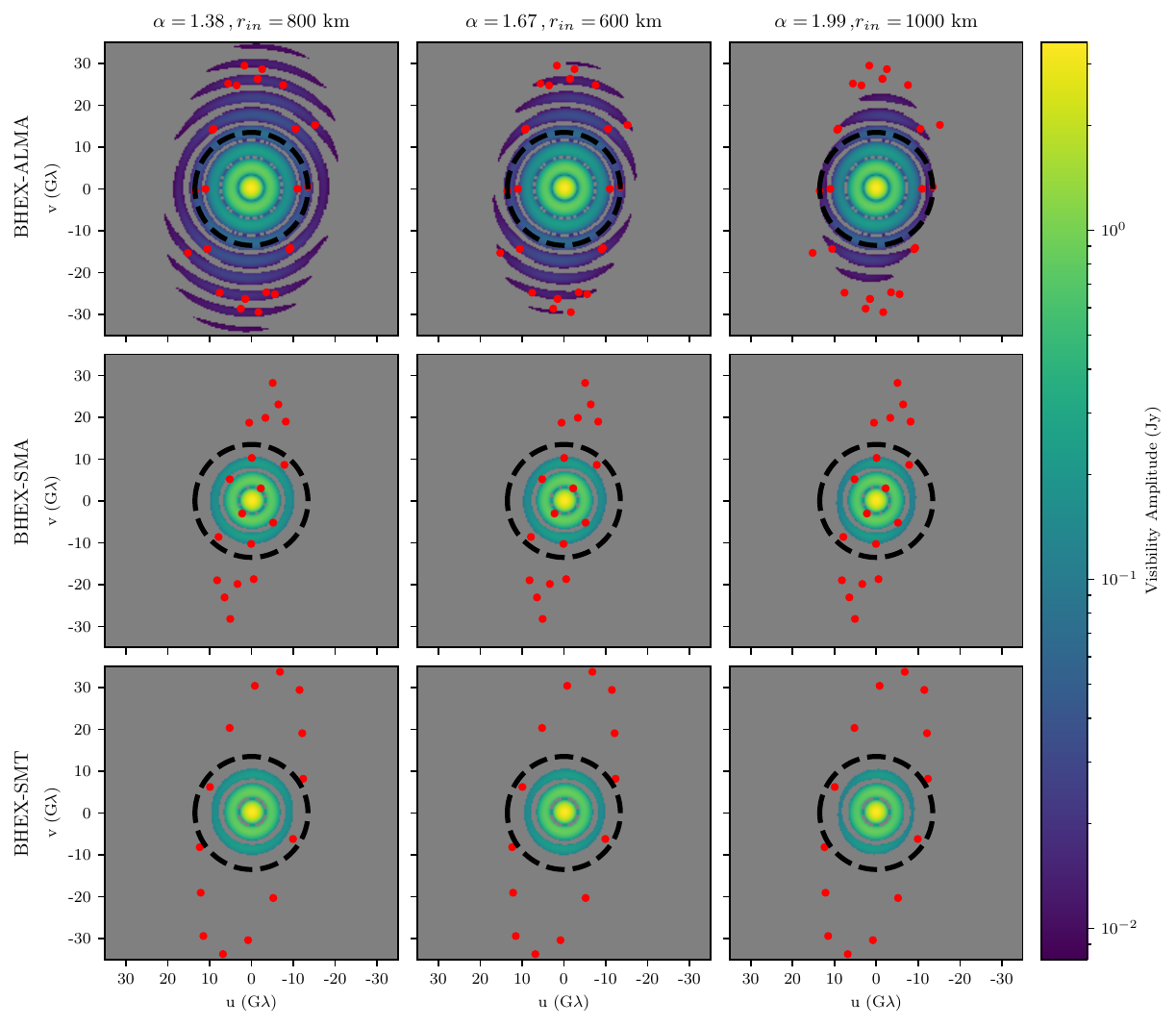}
\caption{The total flux for the BHEX-ALMA, BHEX-SMA and BHEX-SMT baselines for various values of $\alpha$ and $r_{\rm in}$. The colored/gray regions represent the accessible/inaccessible flux, with the floor set by the thermal noise for the corresponding baseline. The red dots represent the $(u,v)$ coverage for the baseline and the black circle is at $13.5$ $G\lambda$ which represents the boundary for our model beyond which the photon ring in linear polarisation is expected to dominate. The colored region represents the flux signal lying above the detection threshold of 3$\sigma$ for each baseline.}
\label{f:sgra_bsl_alphas}
\end{figure*}

\section{Orbit transfers and Earth-space VLBI}\label{sec:orb_transfers}
A space mission has a designated ``Target'' orbit which is chosen to maximise its scientific output. However, it is often the case that the orbiter isn't \textit{directly} injected into that orbit, but is first launched into an initial ``Parking'' orbit from which it performs a series of transfers to go to the Target orbit. As an example, the INTEGRAL mission \citep{Winkler:2003} launched the orbiter first into a low, nearly circular parking orbit, then into a highly elliptical transfer orbit and finally into the target geosynchonous orbit \citep{Eismont:2003,Jensen:2003}.

For this paper, the Target orbit is chosen to be the reference orbit of the BHEX mission \citep{Johnson:2024}. This orbit has been preliminarily selected for BHEX for the following reasons:
\begin{itemize}
    \item The selected altitude provides the required angular resolution on ground-space baselines to probe the photon ring. An orbital period of 12 sidereal hours also generates a repeating ground track of the spacecraft which simplifies selection of ground station locations for data downlink.
    \item The inclination maximises the projected baseline length towards both \s{} and M87*.
    \item For M87*, precise estimates of the mass are not available which requires that mass/spin degeneracies in the photon ring must be broken with two-dimensional information about its shape and relative astrometry. This requires the near-circular \emph{(u,v)} coverage by selecting an orbital plane almost perpendicular to M87*. For \s, knowledge of the mass is available to  a finer degree of precision. This means that a photon ring size measurement along even a single axis provides an excellent spin constraint. The orbital plane has however been rotated slightly to increase the coverage of \s{} in the \emph{u}-plane.
\end{itemize}
\noindent To reach the Target orbit, we propose a  ``transfer-observe-transfer'' orbital migration scheme which starts from a Parking orbit that is in the Low-Earth-Orbit (LEO) region, having a semi-major axis of $a=7000$ km. The orbiter performs 1 day of observations in this orbit. Then, it goes to two Intermediate Circular orbits (hereafter IC1 and IC2) having $a=13,000$ and $a=19,000$ km respectively, observing for 1 day in each of them. We note that an extended time can be spent in the intermediate orbits if there are any unforeseen challenges of observing on a particular day. Finally, the orbiter goes to the Target BHEX orbit and observes for 1 day. Since all the aforementioned orbits are circular, the argument of perigee is an undefined quantity (and taken to be $0^{\circ}$ as per convention) and the true anomaly $\nu$ is a free parameter. The corresponding orbital parameters for all of these stages are given in Table \ref{tab:orbit_param}. In the next section, we discuss the two possible astrodynamical methods that can perform the transfer between these orbits.

\begin{table}
\caption{Orbital parameters for performing an orbit migration from the initial Parking orbit to the final Target orbit.}
    \centering
    \begin{tabular}{|c|c|c|c|c|c|c|}
    \hline
    Orbit Type & a (km)&e&i ($^{\circ}$)&$\Omega\, 
 (^{\circ})$&$\omega\, 
 (^{\circ})$&$\nu\, (^{\circ})$\\
    \hline
    Parking&7,000&0&90&247.7&0&-90.00\\
    IC1&13,000&0&90&247.7&0&128.58\\
    
    IC2&19,000&0&90&247.7&0&-66.63\\
    Target&26,563.88&0&90&247.7&0&179.99\\
    \hline
    \end{tabular}
    \tablefoot{The Target orbit's parameters are chosen to be of the proposed BHEX mission's orbiter. Here IC1 and IC2 represent the two Intermediate Circular orbits.}
    \label{tab:orbit_param}
\end{table}
\subsection{Hohmann transfers}
One of the most well-studied orbit transfers is the Hohmann transfer \citep{Vallado}. When the ratio of the radii of the initial and final orbit involved in an orbital transfer is low, it is the most energy efficient transfer between two circular, co-planar orbits \citep{Prussing:1992}. It utilises two impulsive thrusts: one at the perigee of the initial orbit which launches the orbiter into an elliptical transfer orbit, and the second at the apogee of this orbit which launches the orbiter into the final orbit. Since the thrusts are impulsive, a large $\Delta v$ is required from the thruster in a short span of time and therefore Hohmann transfers are performed using CP.

It is easy to recognise the utility of the Hohmann transfer for a mission like BHEX whose proposed final orbit is circular and the standard implementation of the transfer is between circular orbits. However, it is indeed possible that for non-circular Parking or intermediate orbits, more sophisticated and fuel-efficient optimisation schemes exist. Nevertheless, our motivation for working with the Hohmann transfer is to use it as a tool to demonstrate the construction of a mission architecture that addresses the astrophysical requirement identified earlier of making observations at baselines shorter than the ones obtained from BHEX's final orbit. 

A demonstration of the $(u,v)$ coverage for observing \s{}  based on a series of Hohmann transfers is given in Figure \ref{f:alluv}. This represents the $(u,v)$ coverage for a space orbiter being added to the expected EHT 2025 array (see Table \ref{t:ground+dlink}), with the former supporting an observing bandwidth of 16 GHz and an integration time of $t=10s$ \citep{Johnson:2024}. For the Parking orbit, the time between scans is 900s which ensures filling of the $(u,v)$ plane with the orbiter having a relatively short period ($\approx97$ minutes). Although the Parking Orbit does not access the baseline lengths needed to probe the photon ring signal considered in this paper, it can still be used for checking of various sub-sytems of the orbiter.  For IC1, IC2 and the final BHEX orbit this time is chosen to be one-twelfth of the orbital period which comes out to be approximately 20, 36 and 60 minutes respectively). The top panel represents the putative $(u,v)$ coverage obtained using \texttt{ehtim} while the bottom panel shows the coverage generated using the \texttt{spacevlbi} \footnote{\url{https://github.com/bhudson2/spacevlbi}} tool, which models the impact on \emph{(u,v)} coverage if realistic mission constraints are considered (\cite{hudson_python_2024}). The points in shades of green represent the points lying in the region outside $13.5$ G$\lambda$ which was identified in Section \ref{section:diff_scatt_pr} as the point in the Fourier domain where the LP photon ring signature in LP first begins to dominate.

\subsection{Electric orbit raising}
An alternative to the CP-based Hohmann transfer is the EP-based Electric Orbit Raising (EOR). The primary advantage of EP over CP is the significantly higher specific impulse, but that comes at a cost of increased transfer time and typically a much higher power requirement which can drive the spacecraft power system design. Here, the thrust provided in a \textit{single} burn is significantly lower than CP and hence cannot be used to perform the Hohmann transfer. Moreover, the time taken to go from one orbit to another is significantly longer compared to the Hohmann case. Nevertheless, due to a much higher specific impulse, EP is much more fuel efficient than CP and this can be used to increase the payload mass for a given mission. The relevant formulae to compute the time and fuel required for EOR is given in Section \ref{sec:Transfer_Fuel} of the Appendix.

We note that the $(u,v)$ coverage for both CP and EP would crudely look the same since the observing campaign is being triggered only on the four orbits in Table \ref{tab:orbit_param} and \textit{not} in between them. However, where propulsion \textit{does} come into play is the time taken to reach these orbits and the fuel that would be required to perform these orbital maneuvers. This will be discussed in the next Section.

\section{Impact of practical mission considerations on Earth-space VLBI}\label{sec:practical_VLBI}
We now discuss the impact on Earth-space VLBI observations of two crucial mission design considerations for BHEX, namely maintaining a real-time downlink connection with ground stations, and the choice of propulsion.

\begin{figure*}
\centering
\includegraphics[width=\textwidth,height=8cm]{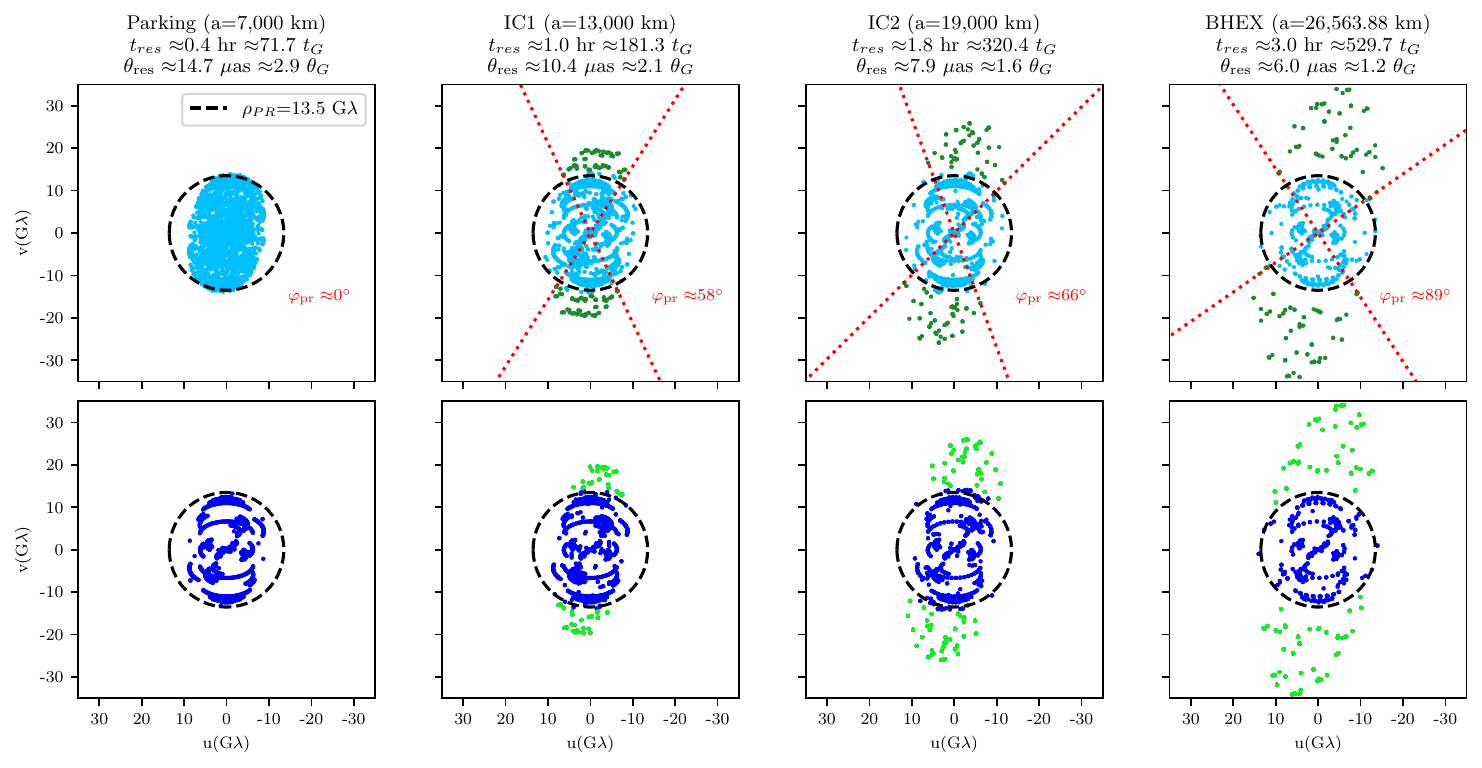}
\caption{The $(u,v)$ coverage of the proposed orbital migration scheme. The temporal and angular resolution is denoted by $t_{res}$ and $\theta_{res}$ respectively with the corresponding values in gravitational units denoted by the subscript $G$. The angle $\varphi_{\rm pr}$ represents the angular sampling of the photon ring at each stage. The top and bottom panel coverage is generated using \texttt{ehtim} and \texttt{spacevlbi} packages respectively, with the latter imposing the constraint of maintaining a real-time downlink connection with ground stations. The VLBI ground array and the downlink station locations are given in Table \ref{t:ground+dlink}}.
\label{f:alluv}
\end{figure*}

\subsection{Maintaining a real-time downlink connection}

In performing VLBI observations from space, there are a number of potential limitations on when observations can be performed, imposed by the spacecraft and wider mission design. These are hereafter called \textit{functional constraints} which in general include (but are not limited to):
\begin{itemize}
    \item Sun, Earth and Moon blinding of star trackers required for high accuracy attitude control,
    \item Radiator surfaces not being deep-space pointed as required for the demanding thermal control of space VLBI missions,
    \item Tracking of the spacecraft for highly accurate orbit determination required for the correlation process.
\end{itemize}
\noindent For BHEX, a major constraint is the preliminary decision to perform real-time downlink of science data to a network of ground stations via an optical link \citep{Wang:2024}. Alternatives to this solution, including the use of mass data storage onboard the spacecraft were evaluated. However, it was determined that the technology readiness of data storage to the level required for BHEX was not sufficiently mature for inclusion in a SMEX mission. Such technologies also come with a significant mass and power requirement. The use of an optical terminal to achieve real-time downlink of the large volumes of data was deemed to be more feasible within the tight constraints of the mission. The difficulties associated with this choice are noted and future work will explore these challenges and propose solutions. 

The impact of this constraint is that observations cannot be performed at times when a link between the spacecraft's optical terminal and the network of ground stations is not possible. Various parameters impact the severity of this constraint including: the attitude control strategy of the spacecraft, the specific ground station configuration, the minimum elevation of the spacecraft at the optical ground stations and the gimbal control limits of the onboard optical terminal.

To study the impact of this constraint on \s's observations, in Figure \ref{f:alluv} the second panel depicts the potential impact on $(u,v)$ coverage caused by the inability to maintain a real-time downlink throughout the spacecraft's orbit. The chosen downlink stations are specified in Table \ref{t:ground+dlink}, with station choices driven by the potential sites identified by BHEX \citep{Wang:2024}. Final selection of the BHEX optical ground station sites is still taking place considering factors such as the orbit coverage, budget and the viability of the site (e.g. weather, technology readiness). This panel, generated using \texttt{spacevlbi}, enables modelling of the major functional constraints impacting a space-based VLBI mission. The implementation in the package includes the effect of Earth's obscuration and the constraint of maintaining a real-time downlink with Earth-based stations. For this example case, an optical terminal with a $\pm 90^{\circ}$ gimbal control mechanism is implemented on the spacecraft. Throughout observations, the science antenna is pointed at \s. The spacecraft's attitude is however rotated about the science antenna direction every half orbit period to keep the optical terminal approximately pointed towards the Earth.

The impact on observations and thus the science return on the mission can be seen in the loss of $(u,v)$ coverage between the top and bottom panels of Figure \ref{f:alluv}. This is more pronounced for lower altitude orbits as four ground stations are not sufficient to provide full coverage of a spacecraft in  LEO. However, we note that a different choice of mean anomaly (which signifies the time at which the orbiter is launched) can lead to a drastically different downlink coverage in the LEO, thereby giving more $(u,v)$ points at this altitude. Indeed, the choice of mean anomaly for this paper was to optimise the coverage with respect to ground for the \textit{final} BHEX orbit, but that has no bearing on the values for the lower orbits. Mitigating the impact of the functional constraints is crucial for maximising the science return of the mission and for BHEX, the real-time downlink is likely to be the dominant factor.

\subsection{Impact of propulsion choices}
To make the propulsion computations for our orbit migration scheme, we make the following choices. The propellant requirements for CP are chosen based on Falcon 9's specifications, which has been NASA's choice for the Small Explorers (SMEX) class of missions like IXPE \citep{Weisskopf:2022} launched in 2021, while EP's expellant parameters are of the SPT-140 Hall thrusters which were recently used by NASA in the Discovery class Psyche mission \citep{Hart:2018,Khan:2024} launched in 2023. We note that BHEX is a proposed SMEX class mission and therefore would use a smaller version of the Psyche propulsion system design. The performance parameters for these propellants along with the detailed computations for the time and fuel requirements for implementing the transfers are given in Section \ref{sec:Transfer_Fuel} of the Appendix.

\subsection{Results and inferences}
We find that assuming a 
spacecraft dry mass of $300$ kg (a typical upper limit on NASA SMEX missions \citep{nasa_smex}, for CP-based Hohmann transfer using cryogenic liquid methane and liquid oxygen (LOX) as a propellant, we would require 195.889 kg of fuel and 9.21 hours to go from the Parking to the Target orbit through our bespoke orbital scheme. For an EP-based EOR using the SPT-140 Hall effect thruster, we would require 57.20 kg of fuel and 44.2 days to complete the migration.

The significantly large amount of fuel required for the Hohmann transfer-based scheme restricts its practical utility for having an orbiter observe in the intermediate stage, at least in the context of the budget constraints of the SMEX mission. The high propellant mass would require a launch vehicle upper stage (like the Falcon 9 Block 5) or a kick stage rocket to perform the transfer. Utilisation of such a system would make it highly unlikely that observations could be performed by BHEX during the transfer as this would require the attitude control system to be designed to accommodate connection to this upper stage, resulting in an over-designed system for the nominal mission.  

On the other hand, the values obtained from EP are encouraging since not only are the fuel requirements significantly lower than the CP case (which is expected), but also the total migration time of ~44 days isn't  prohibitively long with respect to BHEX's 2 year mission \citep{Johnson:2024}, which is often one of the criticisms of using EP. However, it is noted that EP has an extremely high power requirement (the stated SPT-140 system requires 4.5 kW). Utilisation of such a system for BHEX would drive the power system design, resulting in the need for very large solar panels. This may be prohibited by the constraints of a SMEX-class mission. As such, in reality a smaller, lower-thrust EP system would likely be implemented, resulting in a longer transfer time (see Figure \ref{f:EP_time} in Section \ref{sec:Transfer_Fuel} of the Appendix).

\section{Consequences for \s}\label{sec:abc}
Changes in the spacecraft orbit change the temporal and spatial resolution of the full VLBI array containing the orbiter. The spatial resolution element of the array can be estimated in a number of ways \citep{Thompson}, but is generally inversely proportional to the longest baseline and estimated with
\begin{align}
    \theta_{\rm res} &\equiv \frac{\lambda}{{\rm max}(B)},
\end{align}
where $B$ is the physical length of the baseline.

However, the temporal resolution element is less clearly defined. \citet{Palumbo_2019} used filling of the $(u,v)$ to define an imaging-focused temportal resolution. However, in VLBI applications targeting features varying widely in spatial scale, the uniform $(u,v)$ sampling weight is less sensible.

In this paper, we simply define the temporal resolution afforded by the spacecraft to be one quarter of its orbital period. Due to the complex-conjugate symmetry of the $(u,v)$ plane, a dish in a circular orbit samples essentially all of its $(u,v)$ track in half of its orbit, and so a quarter period represents accruing more than half of the image information at the spatial scales targeted by the orbiter:
\begin{align}
    t_{\rm res} \equiv \frac{P}{4}.
\end{align}

For the photon ring morphological science targeted by the BHEX mission, azimuthal angle sampling in the $(u,v)$ plane is also important, as different baseline orientations project out different information about photon ring morphology \citep{Bracewell_1956, GLM_2020}. This angular sampling must occur on baselines not dominated by the direct image structure. To characterize the thoroughness of this sampling over the course of several transfers, we characterize the angular sampling of the photon ring as $\varphi_{\rm pr}$, the widest range of $(u,v)$ angles sampled by a non-conjugate-redundant subset of the u-v coverage beyond the minimum photon ring $(u,v)$ radius, in this case 13.5 G$\lambda$.

Figure \ref{f:alluv} shows three diagnostic properties along with the evolving orbit: the temporal resolution $t_{\rm res}$, the spatial resolution $\theta_{\rm res}$, and the angular sampling beyond the photon ring, $\varphi_{\rm pr}$. 
The Figure also contains values obtained in gravitational units (denoted by the subscript $G$). These require fixing the mass and distance from Earth of \s{} and we consider the values to be $M_{\rm S}=4.1\times10^{6}M_{\odot}$ and $D_{S}=8.1$ kpc respectively \citep{GRAVITY:2018}. Then, the quantities $t_{\rm G}$ and $\theta_{\rm G} $ are obtained by dividing $t_{\rm res}$ and $\theta_{\rm res}$ by $GM_{S}/c^{3}$ and $GM_{S}/c^{2}D_{S}$ respectively, where $G$ is Newton's gravitational constant and $c$ is the speed of light. The quantity $GM_{S}/c^{2}D_{S}$ corresponds to one gravitational radius and serves as a useful unit in constructing quantities sensitive to strong field gravity signatures of \s{} \citep{Psaltis:2015}. From the values obtained for each stage, we note that raising the orbit worsens the temporal resolution, but improves the spatial resolution and angular sampling of the photon ring. 

Each transfer sacrifices more signal to scattering while plumbing finer structures in the source. Depending on the accretion and interstellar scattering conditions in \s, any of the latter three orbits may be sufficient or even optimal for measuring the spin of \s, as lower orbits place coverage in regions with detectable signal in Figure \ref{f:sgra_bsl_alphas}.

\section{Conclusion and future work} \label{sec:conclusions}

In this paper we have studied the extent to which diffractive scattering affects Earth-space VLBI observations of \s{}, focusing primarily on the BHEX mission. A simple geometric model for concentric Gaussian rings was considered for modelling \s{}'s photon ring signature and the effects of diffractive scattering were studied by varying the power-law index $\alpha$ of the phase structure function  and the inner scale for the onset of turbulence $r_{\rm in}$. Through thermal noise calculations, it was shown that for variations of $\alpha$ from 1.38 to 1.99 and $r_{\rm in}$ from 600 to 1000 km, for the most sensitive baseline, namely BHEX-ALMA, the longest baselines of the orbiter will be sensitive to \s{}'s signal. However for baselines with moderate sensitivity such as BHEX-SMA and BHEX-SMT, the longest baselines are unable to probe the signal, with the situation getting particularly worse for higher values of $\alpha$.

These lessons were used to motivate a dynamic orbit migration scheme for the mission wherein instead of just observing in the fixed, final orbit, the orbiter goes through two intermediate orbits and observes in each of them. The $(u,v)$ coverage in these orbits will lie in regions where the sensitivity requirements are not prohibitive to observations, whilst also being robust to variations in $\alpha$ and $r_{\rm in}$.

Some of the main, detailed mission design considerations for this implementation were also studied. We discussed the importance of including the constraint of maintaining a downlink connection with Earth stations and generated the corresponding $(u,v)$ coverage. For the orbit migration, the Hohmann transfer was presented as a viable option. Subsequently, a quantitative study is conducted of the fuel requirements for both CP and EP, focusing on currently used technologies by NASA's missions and associated launch vehicles. Lastly, the utility of these orbits to probe \s's rapid timescales and photon ring signatures is explored using spatial and temporal resolution estimates, as well as angular sampling of the photon ring signal.

The BHEX mission concept currently requires the study of the M87* and \s{} photon rings using the same orbit; these two sources give a trade space for which the Target BHEX orbit is one of many reasonable compromises. However, were \s{} the only target, the Fourier coverage indicates that slightly shorter baselines may be preferred. While a mission window with shorter baselines can in principle be done in a few hours by a CP-based Hohmann transfer scheme using the LOX propellant, the corresponding fuel requirement is prohibitively excessive. On the other hand, EP based EoR using the SPT-140 thrusters provides a viable mechanism of reaching and observing in the Transfer orbits in a fuel efficient manner, requiring comparatively lesser fuel but also reaching the Target BHEX orbit in about 6 weeks. As the science case for BHEX mission is developed, it will be interesting to explore whether any of these considerations can aid in achieving its scientific goals, particularly for observing \s. More broadly, it seems that more probably than not, a hybrid propulsion scheme \citep{Mailhe:2022} might be best suited for future Earth-space VLBI missions primarily focused on studying \s.

In future work, we aim to perform a detailed study of how Earth-space VLBI observations of the Galactic Centre can help address the uncertainties in various physical processes characterising turbulence in the ISM.

\begin{acknowledgements}
The authors are grateful to the anonymous referee, whose feedback played a pivotal role in obtaining the results presented in this paper. We would also like to thank Michael D. Johnson and Sara Issaoun for several fruitful discussions and Leonid Gurvits for providing valuable comments that helped improve this paper. A.T. gratefully acknowledges the Center for Astrophysics | Harvard \& Smithsonian, and the Black Hole Initiative at Harvard University for
providing a stimulating environment during an extended research visit between May-July 2024 during which a part of this work was done.
\end{acknowledgements}

\bibliographystyle{aa} 
\bibliography{aa.bib}

\begin{thebibliography}{93}
\expandafter\ifx\csname natexlab\endcsname\relax\def\natexlab#1{#1}\fi

\bibitem[{{Andrianov} {et~al.}(2021){Andrianov}, {Baryshev}, {Falcke}, {Girin}, {de Graauw}, {Kostenko}, {Kudriashov}, {Ladygin}, {Likhachev}, {Roelofs}, {Rudnitskiy}, {Shaykhutdinov}, {Shchekinov}, \& {Shchurov}}]{Andrianov:2021}
{Andrianov}, A.~S., {Baryshev}, A.~M., {Falcke}, H., {et~al.} 2021, \mnras, 500, 4866

\bibitem[{{Armstrong} {et~al.}(1995){Armstrong}, {Rickett}, \& {Spangler}}]{Armstrong:1995}
{Armstrong}, J.~W., {Rickett}, B.~J., \& {Spangler}, S.~R. 1995, \apj, 443, 209

\bibitem[{Bhat {et~al.}(2004)Bhat, Cordes, Camilo, Nice, \& Lorimer}]{Bhat:2004}
Bhat, N. D.~R., Cordes, J.~M., Camilo, F., Nice, D.~J., \& Lorimer, D.~R. 2004, Astrophys. J., 605, 759

\bibitem[{{Bower} {et~al.}(2015){Bower}, {Markoff}, {Dexter}, {Gurwell}, {Moran}, {Brunthaler}, {Falcke}, {Fragile}, {Maitra}, {Marrone}, {Peck}, {Rushton}, \& {Wright}}]{Bower:2015}
{Bower}, G.~C., {Markoff}, S., {Dexter}, J., {et~al.} 2015, \apj, 802, 69

\bibitem[{{Bracewell}(1956)}]{Bracewell_1956}
{Bracewell}, R.~N. 1956, Australian Journal of Physics, 9, 198

\bibitem[{{Broderick} {et~al.}(2022){Broderick}, {Tiede}, {Pesce}, \& {Gold}}]{Broderick:2022}
{Broderick}, A.~E., {Tiede}, P., {Pesce}, D.~W., \& {Gold}, R. 2022, \apj, 927, 6

\bibitem[{{C{\'a}rdenas-Avenda{\~n}o} \& {Lupsasca}(2023)}]{Avendano:2023}
{C{\'a}rdenas-Avenda{\~n}o}, A. \& {Lupsasca}, A. 2023, \prd, 108, 064043

\bibitem[{{C{\'a}rdenas-Avenda{\~n}o} {et~al.}(2023){C{\'a}rdenas-Avenda{\~n}o}, {Lupsasca}, \& {Zhu}}]{Avendano:2023A}
{C{\'a}rdenas-Avenda{\~n}o}, A., {Lupsasca}, A., \& {Zhu}, H. 2023, \prd, 107, 043030

\bibitem[{Chepurnov \& Lazarian(2010)}]{Chepurnov:2010}
Chepurnov, A. \& Lazarian, A. 2010, Astrophys. J., 710, 853

\bibitem[{{Cho} {et~al.}(2022){Cho}, {Zhao}, {Kawashima}, {Kino}, {Akiyama}, {Johnson}, {Issaoun}, {Moriyama}, {Cheng}, {Algaba}, {Jung}, {Sohn}, {Krichbaum}, {Wielgus}, {Hada}, {Lu}, {Cui}, {Sawada-Satoh}, {Shen}, {Park}, {Jiang}, {Ro}, {Yi}, {Wajima}, {Lee}, {Hodgson}, {Tazaki}, {Honma}, {Niinuma}, {Trippe}, {An}, {Zhang}, {Lee}, {Oh}, {Byun}, {Lee}, {Kim}, {Oh}, {Koyama}, {Asada}, {Wang}, {Cui}, {Hagiwara}, {Nakamura}, {Takamura}, {Hirota}, {Sugiyama}, {Kawaguchi}, {Kobayashi}, {Oyama}, {Yonekura}, {Kim}, {Hwang}, {Jung}, {Kim}, {Kim}, {Oh}, {Roh}, {Yeom}, {Xia}, {Zhong}, {Li}, {Zhao}, {Wang}, {Liu}, \& {Chen}}]{Cho:2022}
{Cho}, I., {Zhao}, G.-Y., {Kawashima}, T., {et~al.} 2022, \apj, 926, 108

\bibitem[{{Cordes} {et~al.}(1986){Cordes}, {Pidwerbetsky}, \& {Lovelace}}]{Cordes:1986}
{Cordes}, J.~M., {Pidwerbetsky}, A., \& {Lovelace}, R.~V.~E. 1986, \apj, 310, 737

\bibitem[{{Cordes} {et~al.}(1985){Cordes}, {Weisberg}, \& {Boriakoff}}]{Cordes:1985}
{Cordes}, J.~M., {Weisberg}, J.~M., \& {Boriakoff}, V. 1985, \apj, 288, 221

\bibitem[{{Cordes} \& {Wolszczan}(1986)}]{Cordes:1986A}
{Cordes}, J.~M. \& {Wolszczan}, A. 1986, \apjl, 307, L27

\bibitem[{{Davies} {et~al.}(1976){Davies}, {Walsh}, \& {Booth}}]{Davies:1976}
{Davies}, R.~D., {Walsh}, D., \& {Booth}, R.~S. 1976, \mnras, 177, 319

\bibitem[{{Doeleman} {et~al.}(2023){Doeleman}, {Barrett}, {Blackburn}, {Bouman}, {Broderick}, {Chaves}, {Fish}, {Fitzpatrick}, {Freeman}, {Fuentes}, {G{\'o}mez}, {Haworth}, {Houston}, {Issaoun}, {Johnson}, {Kettenis}, {Loinard}, {Nagar}, {Narayanan}, {Oppenheimer}, {Palumbo}, {Patel}, {Pesce}, {Raymond}, {Roelofs}, {Srinivasan}, {Tiede}, {Weintroub}, \& {Wielgus}}]{Doeleman:2023}
{Doeleman}, S.~S., {Barrett}, J., {Blackburn}, L., {et~al.} 2023, Galaxies, 11, 107

\bibitem[{{Eismont} {et~al.}(2003){Eismont}, {Ditrikh}, {Janin}, {Karrask}, {Clausen}, {Medvedchikov}, {Kulik}, {Vtorushin}, \& {Yakushin}}]{Eismont:2003}
{Eismont}, N.~A., {Ditrikh}, A.~V., {Janin}, G., {et~al.} 2003, \aap, 411, L37

\bibitem[{{Event Horizon Telescope Collaboration} {et~al.}(2024{\natexlab{a}}){Event Horizon Telescope Collaboration}, {Akiyama}, {Alberdi}, {Alef}, {Algaba}, {Anantua}, {Asada}, {Azulay}, {Bach}, {Baczko}, {Ball}, {Balokovic}, {Bandyopadhyay}, {Barrett}, {Baub{\"o}ck}, {Benson}, {Bintley}, {Blackburn}, {Blundell}, {Bouman}, {Bower}, {Boyce}, {Bremer}, {Brinkerink}, {Brissenden}, {Britzen}, {Broderick}, {Broguiere}, {Bronzwaer}, {Bustamante}, {Byun}, {Carlstrom}, {Ceccobello}, {Chael}, {Chan}, {Chang}, {Chatterjee}, {Chatterjee}, {Chen}, {Chen}, {Cheng}, {Cho}, {Christian}, {Conroy}, {Conway}, {Cordes}, {Crawford}, {Crew}, {Cruz-Osorio}, {Cui}, {Dahale}, {Davelaar}, {De Laurentis}, {Deane}, {Dempsey}, {Desvignes}, {Dexter}, {Dhruv}, {Dihingia}, {Doeleman}, {Dougal}, {Dzib}, {Eatough}, {Emami}, {Falcke}, {Farah}, {Fish}, {Fomalont}, {Ford}, {Foschi}, {Fraga-Encinas}, {Freeman}, {Friberg}, {Fromm}, {Fuentes}, {Galison}, {Gammie}, {Garc{\'\i}a}, {Gentaz}, {Georgiev}, {Goddi}, {Gold}, {G{\'o}mez-Ruiz},
  {G{\'o}mez}, {Gu}, {Gurwell}, {Hada}, {Haggard}, {Haworth}, {Hecht}, {Hesper}, {Heumann}, {Ho}, {Ho}, {Honma}, {Huang}, {Huang}, {Hughes}, {Ikeda}, {Impellizzeri}, {Inoue}, {Issaoun}, {James}, {Jannuzi}, {Janssen}, {Jeter}, {Jiang}, {Jim{\'e}nez-Rosales}, {Johnson}, {Jorstad}, {Joshi}, {Jung}, {Karami}, {Karuppusamy}, {Kawashima}, {Keating}, {Kettenis}, {Kim}, {Kim}, {Kim}, {Kim}, {Kino}, {Koay}, {Kocherlakota}, {Kofuji}, {Koch}, {Koyama}, {Kramer}, {Kramer}, {Kramer}, {Krichbaum}, {Kuo}, {La Bella}, {Lauer}, {Lee}, {Lee}, {Leung}, {Levis}, {Li}, {Lico}, {Lindahl}, {Lindqvist}, {Lisakov}, {Liu}, {Liu}, {Liuzzo}, {Lo}, {Lobanov}, {Loinard}, {Lonsdale}, {Lowitz}, {Lu}, {MacDonald}, {Mao}, {Marchili}, {Markoff}, {Marrone}, {Marscher}, {Mart{\'\i}-Vidal}, {Matsushita}, {Matthews}, {Medeiros}, {Menten}, {Michalik}, {Mizuno}, {Mizuno}, {Moran}, {Moriyama}, {Moscibrodzka}, {Mulaudzi}, {M{\"u}ller}, {M{\"u}ller}, {Mus}, {Musoke}, {Myserlis}, {Nadolski}, {Nagai}, {Nagar}, {Nakamura}, {Narayanan}, {Natarajan},
  {Nathanail}, {Fuentes}, {Neilsen}, {Neri}, {Ni}, {Noutsos}, {Nowak}, {Oh}, {Okino}, {Olivares}, {Ortiz-Le{\'o}n}, {Oyama}, {{\"O}zel}, {Palumbo}, {Paraschos}, {Park}, {Parsons}, {Patel}, {Pen}, {Pesce}, {Pi{\'e}tu}, {Plambeck}, {PopStefanija}, {Porth}, {P{\"o}tzl}, {Prather}, {Preciado-L{\'o}pez}, {Psaltis}, {Pu}, {Ramakrishnan}, {Rao}, {Rawlings}, {Raymond}, {Rezzolla}, {Ricarte}, {Ripperda}, {Roelofs}, {Rogers}, {Romero-Ca{\~n}izales}, {Ros}, {Roshanineshat}, {Rottmann}, {Roy}, {Ruiz}, {Ruszczyk}, {Rygl}, {S{\'a}nchez}, {S{\'a}nchez-Arg{\"u}elles}, {S{\'a}nchez-Portal}, {Sasada}, {Satapathy}, {Savolainen}, {Schloerb}, {Schonfeld}, {Schuster}, {Shao}, {Shen}, {Small}, {Sohn}, {SooHoo}, {Sosapanta Salas}, {Souccar}, {Stanway}, {Sun}, {Tazaki}, {Tetarenko}, {Tiede}, {Tilanus}, {Titus}, {Torne}, {Toscano}, {Traianou}, {Trent}, {Trippe}, {Turk}, {van Bemmel}, {van Langevelde}, {van Rossum}, {Vos}, {Wagner}, {Ward-Thompson}, {Wardle}, {Washington}, {Weintroub}, {Wharton}, {Wielgus}, {Wiik}, {Witzel}, {Wondrak},
  {Wong}, {Wu}, {Yadlapalli}, {Yamaguchi}, {Yfantis}, {Yoon}, {Young}, {Young}, {Younsi}, {Yu}, {Yuan}, {Yuan}, {Zensus}, {Zhang}, {Zhao}, \& {Zhao}}]{EHT_SgrA_VII:2024}
{Event Horizon Telescope Collaboration}, {Akiyama}, K., {Alberdi}, A., {et~al.} 2024{\natexlab{a}}, \apjl, 964, L25

\bibitem[{{Event Horizon Telescope Collaboration} {et~al.}(2024{\natexlab{b}}){Event Horizon Telescope Collaboration}, {Akiyama}, {Alberdi}, {Alef}, {Algaba}, {Anantua}, {Asada}, {Azulay}, {Bach}, {Baczko}, {Ball}, {Balokovi{\'c}}, {Bandyopadhyay}, {Barrett}, {Baub{\"o}ck}, {Benson}, {Bintley}, {Blackburn}, {Blundell}, {Bouman}, {Bower}, {Boyce}, {Bremer}, {Brissenden}, {Britzen}, {Broderick}, {Broguiere}, {Bronzwaer}, {Bustamante}, {Carlstrom}, {Chael}, {Chan}, {Chang}, {Chatterjee}, {Chatterjee}, {Chen}, {Chen}, {Cheng}, {Cho}, {Christian}, {Conroy}, {Conway}, {Crawford}, {Crew}, {Cruz-Osorio}, {Cui}, {Dahale}, {Davelaar}, {De Laurentis}, {Deane}, {Dempsey}, {Desvignes}, {Dexter}, {Dhruv}, {Dihingia}, {Doeleman}, {Dzib}, {Eatough}, {Emami}, {Falcke}, {Farah}, {Fish}, {Fomalont}, {Ford}, {Foschi}, {Fraga-Encinas}, {Freeman}, {Friberg}, {Fromm}, {Fuentes}, {Galison}, {Gammie}, {Garc{\'\i}a}, {Gentaz}, {Georgiev}, {Goddi}, {Gold}, {G{\'o}mez-Ruiz}, {G{\'o}mez}, {Gu}, {Gurwell}, {Hada}, {Haggard}, {Hesper},
  {Heumann}, {Ho}, {Ho}, {Honma}, {Huang}, {Huang}, {Hughes}, {Ikeda}, {Violette Impellizzeri}, {Inoue}, {Issaoun}, {James}, {Jannuzi}, {Janssen}, {Jeter}, {Jiang}, {Jim{\'e}nez-Rosales}, {Johnson}, {Jorstad}, {Jones}, {Joshi}, {Jung}, {Karuppusamy}, {Kawashima}, {Keating}, {Kettenis}, {Kim}, {Kim}, {Kim}, {Kim}, {Kino}, {Koay}, {Kocherlakota}, {Kofuji}, {Koch}, {Koyama}, {Kramer}, {Kramer}, {Kramer}, {Krichbaum}, {Kuo}, {La Bella}, {Lee}, {Levis}, {Li}, {Lico}, {Lindahl}, {Lindqvist}, {Lisakov}, {Liu}, {Liu}, {Liuzzo}, {Lo}, {Lobanov}, {Loinard}, {Lonsdale}, {Lowitz}, {Lu}, {MacDonald}, {Mao}, {Marchili}, {Markoff}, {Marrone}, {Marscher}, {Mart{\'\i}-Vidal}, {Matsushita}, {Matthews}, {Medeiros}, {Menten}, {Mizuno}, {Mizuno}, {Montgomery}, {Moran}, {Moriyama}, {Moscibrodzka}, {Mulaudzi}, {M{\"u}ller}, {M{\"u}ller}, {Mus}, {Musoke}, {Myserlis}, {Nagai}, {Nagar}, {Nakamura}, {Narayanan}, {Natarajan}, {Nathanail}, {Fuentes}, {Neilsen}, {Ni}, {Nowak}, {Oh}, {Okino}, {Olivares}, {Oyama}, {{\"O}zel}, {Palumbo},
  {Paraschos}, {Park}, {Parsons}, {Patel}, {Pen}, {Pesce}, {Pi{\'e}tu}, {PopStefanija}, {Porth}, {Prather}, {Psaltis}, {Pu}, {Ramakrishnan}, {Rao}, {Rawlings}, {Raymond}, {Rezzolla}, {Ricarte}, {Ripperda}, {Roelofs}, {Romero-Ca{\~n}izales}, {Ros}, {Roshanineshat}, {Rottmann}, {Roy}, {Ruiz}, {Ruszczyk}, {Rygl}, {S{\'a}nchez}, {S{\'a}nchez-Arg{\"u}elles}, {S{\'a}nchez-Portal}, {Sasada}, {Satapathy}, {Savolainen}, {Schloerb}, {Schonfeld}, {Schuster}, {Shao}, {Shen}, {Small}, {Sohn}, {SooHoo}, {Salas}, {Souccar}, {Stanway}, {Sun}, {Tazaki}, {Tetarenko}, {Tiede}, {Tilanus}, {Titus}, {Toma}, {Torne}, {Toscano}, {Traianou}, {Trent}, {Trippe}, {Turk}, {van Bemmel}, {van Langevelde}, {van Rossum}, {Vos}, {Wagner}, {Ward-Thompson}, {Wardle}, {Washington}, {Weintroub}, {Wharton}, {Wielgus}, {Wiik}, {Witzel}, {Wondrak}, {Wong}, {Wu}, {Yadlapalli}, {Yamaguchi}, {Yfantis}, {Yoon}, {Young}, {Younsi}, {Yu}, {Yuan}, {Yuan}, {Anton Zensus}, {Zhang}, {Zhao}, {Zhao}, {Allardi}, {Chang}, {Chang}, {Chang}, {Chen}, {Chilson},
  {Faber}, {Gale}, {Han}, {Han}, {Hasegawa}, {Hern{\'a}ndez-Rebollar}, {Huang}, {Jiang}, {Jinchi}, {Kimura}, {Kubo}, {Li}, {Lin}, {Liu}, {Liu}, {Lu}, {Martin-Cocher}, {Meyer-Zhao}, {Monta{\~n}a}, {Moraghan}, {Moreno-Nolasco}, {Nishioka}, {Norton}, {Nystrom}, {Ogawa}, {Oshiro}, {Pradel}, {Principe}, {Raffin}, {Rodr{\'\i}guez-Montoya}, {Shaw}, {Snow}, {Sridharan}, {Srinivasan}, {Wei}, \& {Yu}}]{EHT:2024}
{Event Horizon Telescope Collaboration}, {Akiyama}, K., {Alberdi}, A., {et~al.} 2024{\natexlab{b}}, \aap, 681, A79

\bibitem[{{Event Horizon Telescope Collaboration} {et~al.}(2022{\natexlab{a}}){Event Horizon Telescope Collaboration}, {Akiyama}, {Alberdi}, {Alef}, {Algaba}, {Anantua}, {Asada}, {Azulay}, {Bach}, {Baczko}, {Ball}, {Balokovi{\'c}}, {Barrett}, {Baub{\"o}ck}, {Benson}, {Bintley}, {Blackburn}, {Blundell}, {Bouman}, {Bower}, {Boyce}, {Bremer}, {Brinkerink}, {Brissenden}, {Britzen}, {Broderick}, {Broguiere}, {Bronzwaer}, {Bustamante}, {Byun}, {Carlstrom}, {Ceccobello}, {Chael}, {Chan}, {Chatterjee}, {Chatterjee}, {Chen}, {Chen}, {Cheng}, {Cho}, {Christian}, {Conroy}, {Conway}, {Cordes}, {Crawford}, {Crew}, {Cruz-Osorio}, {Cui}, {Davelaar}, {De Laurentis}, {Deane}, {Dempsey}, {Desvignes}, {Dexter}, {Dhruv}, {Doeleman}, {Dougal}, {Dzib}, {Eatough}, {Emami}, {Falcke}, {Farah}, {Fish}, {Fomalont}, {Ford}, {Fraga-Encinas}, {Freeman}, {Friberg}, {Fromm}, {Fuentes}, {Galison}, {Gammie}, {Garc{\'\i}a}, {Gentaz}, {Georgiev}, {Goddi}, {Gold}, {G{\'o}mez-Ruiz}, {G{\'o}mez}, {Gu}, {Gurwell}, {Hada}, {Haggard}, {Haworth},
  {Hecht}, {Hesper}, {Heumann}, {Ho}, {Ho}, {Honma}, {Huang}, {Huang}, {Hughes}, {Ikeda}, {Impellizzeri}, {Inoue}, {Issaoun}, {James}, {Jannuzi}, {Janssen}, {Jeter}, {Jiang}, {Jim{\'e}nez-Rosales}, {Johnson}, {Jorstad}, {Joshi}, {Jung}, {Karami}, {Karuppusamy}, {Kawashima}, {Keating}, {Kettenis}, {Kim}, {Kim}, {Kim}, {Kim}, {Kino}, {Koay}, {Kocherlakota}, {Kofuji}, {Koch}, {Koyama}, {Kramer}, {Kramer}, {Krichbaum}, {Kuo}, {La Bella}, {Lauer}, {Lee}, {Lee}, {Leung}, {Levis}, {Li}, {Lico}, {Lindahl}, {Lindqvist}, {Lisakov}, {Liu}, {Liu}, {Liuzzo}, {Lo}, {Lobanov}, {Loinard}, {Lonsdale}, {Lu}, {Mao}, {Marchili}, {Markoff}, {Marrone}, {Marscher}, {Mart{\'\i}-Vidal}, {Matsushita}, {Matthews}, {Medeiros}, {Menten}, {Michalik}, {Mizuno}, {Mizuno}, {Moran}, {Moriyama}, {Moscibrodzka}, {M{\"u}ller}, {Mus}, {Musoke}, {Myserlis}, {Nadolski}, {Nagai}, {Nagar}, {Nakamura}, {Narayan}, {Narayanan}, {Natarajan}, {Nathanail}, {Fuentes}, {Neilsen}, {Neri}, {Ni}, {Noutsos}, {Nowak}, {Oh}, {Okino}, {Olivares}, {Ortiz-Le{\'o}n},
  {Oyama}, {{\"O}zel}, {Palumbo}, {Paraschos}, {Park}, {Parsons}, {Patel}, {Pen}, {Pesce}, {Pi{\'e}tu}, {Plambeck}, {PopStefanija}, {Porth}, {P{\"o}tzl}, {Prather}, {Preciado-L{\'o}pez}, {Psaltis}, {Pu}, {Ramakrishnan}, {Rao}, {Rawlings}, {Raymond}, {Rezzolla}, {Ricarte}, {Ripperda}, {Roelofs}, {Rogers}, {Ros}, {Romero-Ca{\~n}izales}, {Roshanineshat}, {Rottmann}, {Roy}, {Ruiz}, {Ruszczyk}, {Rygl}, {S{\'a}nchez}, {S{\'a}nchez-Arg{\"u}elles}, {S{\'a}nchez-Portal}, {Sasada}, {Satapathy}, {Savolainen}, {Schloerb}, {Schonfeld}, {Schuster}, {Shao}, {Shen}, {Small}, {Sohn}, {SooHoo}, {Souccar}, {Sun}, {Tazaki}, {Tetarenko}, {Tiede}, {Tilanus}, {Titus}, {Torne}, {Traianou}, {Trent}, {Trippe}, {Turk}, {van Bemmel}, {van Langevelde}, {van Rossum}, {Vos}, {Wagner}, {Ward-Thompson}, {Wardle}, {Weintroub}, {Wex}, {Wharton}, {Wielgus}, {Wiik}, {Witzel}, {Wondrak}, {Wong}, {Wu}, {Yamaguchi}, {Yoon}, {Young}, {Young}, {Younsi}, {Yuan}, {Yuan}, {Zensus}, {Zhang}, {Zhao}, {Zhao}, {Agurto}, {Allardi}, {Amestica}, {Araneda},
  {Arriagada}, {Berghuis}, {Bertarini}, {Berthold}, {Blanchard}, {Brown}, {C{\'a}rdenas}, {Cantzler}, {Caro}, {Castillo-Dom{\'\i}nguez}, {Chan}, {Chang}, {Chang}, {Chang}, {Chang}, {Chen}, {Chilson}, {Chuter}, {Ciechanowicz}, {Colin-Beltran}, {Coulson}, {Crowley}, {Degenaar}, {Dornbusch}, {Dur{\'a}n}, {Everett}, {Faber}, {Forster}, {Fuchs}, {Gale}, {Geertsema}, {Gonz{\'a}lez}, {Graham}, {Gueth}, {Halverson}, {Han}, {Han}, {Hasegawa}, {Hern{\'a}ndez-Rebollar}, {Herrera}, {Herrero-Illana}, {Heyminck}, {Hirota}, {Hoge}, {Hostler Schimpf}, {Howie}, {Huang}, {Jiang}, {Jinchi}, {John}, {Kimura}, {Klein}, {Kubo}, {Kuroda}, {Kwon}, {Lacasse}, {Laing}, {Leitch}, {Li}, {Liu}, {Liu}, {Lin}, {Lu}, {Mac-Auliffe}, {Martin-Cocher}, {Matulonis}, {Maute}, {Messias}, {Meyer-Zhao}, {Monta{\~n}a}, {Montenegro-Montes}, {Montgomerie}, {Moreno Nolasco}, {Muders}, {Nishioka}, {Norton}, {Nystrom}, {Ogawa}, {Olivares}, {Oshiro}, {P{\'e}rez-Beaupuits}, {Parra}, {Phillips}, {Poirier}, {Pradel}, {Qiu}, {Raffin}, {Rahlin}, {Ram{\'\i}rez},
  {Ressler}, {Reynolds}, {Rodr{\'\i}guez-Montoya}, {Saez-Madain}, {Santana}, {Shaw}, {Shirkey}, {Silva}, {Snow}, {Sousa}, {Sridharan}, {Stahm}, {Stark}, {Test}, {Torstensson}, {Venegas}, {Walther}, {Wei}, {White}, {Wieching}, {Wijnands}, {Wouterloot}, {Yu}, {Yu (于威)}, \& {Zeballos}}]{EHT_Sgr_I}
{Event Horizon Telescope Collaboration}, {Akiyama}, K., {Alberdi}, A., {et~al.} 2022{\natexlab{a}}, \apjl, 930, L12

\bibitem[{{Event Horizon Telescope Collaboration} {et~al.}(2022{\natexlab{b}}){Event Horizon Telescope Collaboration}, {Akiyama}, {Alberdi}, {Alef}, {Algaba}, {Anantua}, {Asada}, {Azulay}, {Bach}, {Baczko}, {Ball}, {Balokovi{\'c}}, {Barrett}, {Baub{\"o}ck}, {Benson}, {Bintley}, {Blackburn}, {Blundell}, {Bouman}, {Bower}, {Boyce}, {Bremer}, {Brinkerink}, {Brissenden}, {Britzen}, {Broderick}, {Broguiere}, {Bronzwaer}, {Bustamante}, {Byun}, {Carlstrom}, {Ceccobello}, {Chael}, {Chan}, {Chatterjee}, {Chatterjee}, {Chen}, {Chen}, {Cheng}, {Cho}, {Christian}, {Conroy}, {Conway}, {Cordes}, {Crawford}, {Crew}, {Cruz-Osorio}, {Cui}, {Davelaar}, {De Laurentis}, {Deane}, {Dempsey}, {Desvignes}, {Dexter}, {Dhruv}, {Doeleman}, {Dougal}, {Dzib}, {Eatough}, {Emami}, {Falcke}, {Farah}, {Fish}, {Fomalont}, {Ford}, {Fraga-Encinas}, {Freeman}, {Friberg}, {Fromm}, {Fuentes}, {Galison}, {Gammie}, {Garc{\'\i}a}, {Gentaz}, {Georgiev}, {Goddi}, {Gold}, {G{\'o}mez-Ruiz}, {G{\'o}mez}, {Gu}, {Gurwell}, {Hada}, {Haggard}, {Haworth},
  {Hecht}, {Hesper}, {Heumann}, {Ho}, {Ho}, {Honma}, {Huang}, {Huang}, {Hughes}, {Ikeda}, {Impellizzeri}, {Inoue}, {Issaoun}, {James}, {Jannuzi}, {Janssen}, {Jeter}, {Jiang}, {Jim{\'e}nez-Rosales}, {Johnson}, {Jorstad}, {Joshi}, {Jung}, {Karami}, {Karuppusamy}, {Kawashima}, {Keating}, {Kettenis}, {Kim}, {Kim}, {Kim}, {Kim}, {Kino}, {Koay}, {Kocherlakota}, {Kofuji}, {Koch}, {Koyama}, {Kramer}, {Kramer}, {Krichbaum}, {Kuo}, {La Bella}, {Lauer}, {Lee}, {Lee}, {Leung}, {Levis}, {Li}, {Lico}, {Lindahl}, {Lindqvist}, {Lisakov}, {Liu}, {Liu}, {Liuzzo}, {Lo}, {Lobanov}, {Loinard}, {Lonsdale}, {Lu}, {Mao}, {Marchili}, {Markoff}, {Marrone}, {Marscher}, {Mart{\'\i}-Vidal}, {Matsushita}, {Matthews}, {Medeiros}, {Menten}, {Michalik}, {Mizuno}, {Mizuno}, {Moran}, {Moriyama}, {Moscibrodzka}, {M{\"u}ller}, {Mus}, {Musoke}, {Myserlis}, {Nadolski}, {Nagai}, {Nagar}, {Nakamura}, {Narayan}, {Narayanan}, {Natarajan}, {Nathanail}, {Fuentes}, {Neilsen}, {Neri}, {Ni}, {Noutsos}, {Nowak}, {Oh}, {Okino}, {Olivares}, {Ortiz-Le{\'o}n},
  {Oyama}, {{\"O}zel}, {Palumbo}, {Paraschos}, {Park}, {Parsons}, {Patel}, {Pen}, {Pesce}, {Pi{\'e}tu}, {Plambeck}, {PopStefanija}, {Porth}, {P{\"o}tzl}, {Prather}, {Preciado-L{\'o}pez}, \& {Psaltis}}]{EHT:SgrA_III}
{Event Horizon Telescope Collaboration}, {Akiyama}, K., {Alberdi}, A., {et~al.} 2022{\natexlab{b}}, \apjl, 930, L14

\bibitem[{{Event Horizon Telescope Collaboration} {et~al.}(2022{\natexlab{c}}){Event Horizon Telescope Collaboration}, {Akiyama}, {Alberdi}, {Alef}, {Algaba}, {Anantua}, {Asada}, {Azulay}, {Bach}, {Baczko}, {Ball}, {Balokovi{\'c}}, {Barrett}, {Baub{\"o}ck}, {Benson}, {Bintley}, {Blackburn}, {Blundell}, {Bouman}, {Bower}, {Boyce}, {Bremer}, {Brinkerink}, {Brissenden}, {Britzen}, {Broderick}, {Broguiere}, {Bronzwaer}, {Bustamante}, {Byun}, {Carlstrom}, {Ceccobello}, {Chael}, {Chan}, {Chatterjee}, {Chatterjee}, {Chen}, {Chen}, {Cheng}, {Cho}, {Christian}, {Conroy}, {Conway}, {Cordes}, {Crawford}, {Crew}, {Cruz-Osorio}, {Cui}, {Davelaar}, {De Laurentis}, {Deane}, {Dempsey}, {Desvignes}, {Dexter}, {Dhruv}, {Doeleman}, {Dougal}, {Dzib}, {Eatough}, {Emami}, {Falcke}, {Farah}, {Fish}, {Fomalont}, {Ford}, {Fraga-Encinas}, {Freeman}, {Friberg}, {Fromm}, {Fuentes}, {Galison}, {Gammie}, {Garc{\'\i}a}, {Gentaz}, {Georgiev}, {Goddi}, {Gold}, {G{\'o}mez-Ruiz}, {G{\'o}mez}, {Gu}, {Gurwell}, {Hada}, {Haggard}, {Haworth},
  {Hecht}, {Hesper}, {Heumann}, {Ho}, {Ho}, {Honma}, {Huang}, {Huang}, {Hughes}, {Ikeda}, {Impellizzeri}, {Inoue}, {Issaoun}, {James}, {Jannuzi}, {Janssen}, {Jeter}, {Jiang}, {Jim{\'e}nez-Rosales}, {Johnson}, {Jorstad}, {Joshi}, {Jung}, {Karami}, {Karuppusamy}, {Kawashima}, {Keating}, {Kettenis}, {Kim}, {Kim}, {Kim}, {Kim}, {Kino}, {Koay}, {Kocherlakota}, {Kofuji}, {Koch}, {Koyama}, {Kramer}, {Kramer}, {Krichbaum}, {Kuo}, {La Bella}, {Lauer}, {Lee}, {Lee}, {Leung}, {Levis}, {Li}, {Lico}, {Lindahl}, {Lindqvist}, {Lisakov}, {Liu}, {Liu}, {Liuzzo}, {Lo}, {Lobanov}, {Loinard}, {Lonsdale}, {Lu}, {Mao}, {Marchili}, {Markoff}, {Marrone}, {Marscher}, {Mart{\'\i}-Vidal}, {Matsushita}, {Matthews}, {Medeiros}, {Menten}, {Michalik}, {Mizuno}, {Mizuno}, {Moran}, {Moriyama}, {Moscibrodzka}, {M{\"u}ller}, {Mus}, {Musoke}, {Myserlis}, {Nadolski}, {Nagai}, {Nagar}, {Nakamura}, {Narayan}, {Narayanan}, {Natarajan}, {Nathanail}, {Fuentes}, {Neilsen}, {Neri}, {Ni}, {Noutsos}, {Nowak}, {Oh}, {Okino}, {Olivares}, {Ortiz-Le{\'o}n},
  {Oyama}, {{\"O}zel}, {Palumbo}, {Paraschos}, {Park}, {Parsons}, {Patel}, {Pen}, {Pesce}, {Pi{\'e}tu}, {Plambeck}, {PopStefanija}, {Porth}, {P{\"o}tzl}, {Prather}, {Preciado-L{\'o}pez}, \& {Psaltis}}]{EHT:SgrA_I}
{Event Horizon Telescope Collaboration}, {Akiyama}, K., {Alberdi}, A., {et~al.} 2022{\natexlab{c}}, \apjl, 930, L12

\bibitem[{{Event Horizon Telescope Collaboration} {et~al.}(2019{\natexlab{a}}){Event Horizon Telescope Collaboration}, {Akiyama}, {Alberdi}, {Alef}, {Asada}, {Azulay}, {Baczko}, {Ball}, {Balokovi{\'c}}, {Barrett}, {Bintley}, {Blackburn}, {Boland}, {Bouman}, {Bower}, {Bremer}, {Brinkerink}, {Brissenden}, {Britzen}, {Broderick}, {Broguiere}, {Bronzwaer}, {Byun}, {Carlstrom}, {Chael}, {Chan}, {Chatterjee}, {Chatterjee}, {Chen}, {Chen}, {Cho}, {Christian}, {Conway}, {Cordes}, {Crew}, {Cui}, {Davelaar}, {De Laurentis}, {Deane}, {Dempsey}, {Desvignes}, {Dexter}, {Doeleman}, {Eatough}, {Falcke}, {Fish}, {Fomalont}, {Fraga-Encinas}, {Freeman}, {Friberg}, {Fromm}, {G{\'o}mez}, {Galison}, {Gammie}, {Garc{\'\i}a}, {Gentaz}, {Georgiev}, {Goddi}, {Gold}, {Gu}, {Gurwell}, {Hada}, {Hecht}, {Hesper}, {Ho}, {Ho}, {Honma}, {Huang}, {Huang}, {Hughes}, {Ikeda}, {Inoue}, {Issaoun}, {James}, {Jannuzi}, {Janssen}, {Jeter}, {Jiang}, {Johnson}, {Jorstad}, {Jung}, {Karami}, {Karuppusamy}, {Kawashima}, {Keating}, {Kettenis}, {Kim},
  {Kim}, {Kim}, {Kino}, {Koay}, {Koch}, {Koyama}, {Kramer}, {Kramer}, {Krichbaum}, {Kuo}, {Lauer}, {Lee}, {Li}, {Li}, {Lindqvist}, {Liu}, {Liuzzo}, {Lo}, {Lobanov}, {Loinard}, {Lonsdale}, {Lu}, {MacDonald}, {Mao}, {Markoff}, {Marrone}, {Marscher}, {Mart{\'\i}-Vidal}, {Matsushita}, {Matthews}, {Medeiros}, {Menten}, {Mizuno}, {Mizuno}, {Moran}, {Moriyama}, {Moscibrodzka}, {M{\"u}ller}, {Nagai}, {Nagar}, {Nakamura}, {Narayan}, {Narayanan}, {Natarajan}, {Neri}, {Ni}, {Noutsos}, {Okino}, {Olivares}, {Ortiz-Le{\'o}n}, {Oyama}, {{\"O}zel}, {Palumbo}, {Patel}, {Pen}, {Pesce}, {Pi{\'e}tu}, {Plambeck}, {PopStefanija}, {Porth}, {Prather}, {Preciado-L{\'o}pez}, {Psaltis}, {Pu}, {Ramakrishnan}, {Rao}, {Rawlings}, {Raymond}, {Rezzolla}, {Ripperda}, {Roelofs}, {Rogers}, {Ros}, {Rose}, {Roshanineshat}, {Rottmann}, {Roy}, {Ruszczyk}, {Ryan}, {Rygl}, {S{\'a}nchez}, {S{\'a}nchez-Arguelles}, {Sasada}, {Savolainen}, {Schloerb}, {Schuster}, {Shao}, {Shen}, {Small}, {Sohn}, {SooHoo}, {Tazaki}, {Tiede}, {Tilanus}, {Titus}, {Toma},
  {Torne}, {Trent}, {Trippe}, {Tsuda}, {van Bemmel}, {van Langevelde}, {van Rossum}, {Wagner}, {Wardle}, {Weintroub}, {Wex}, {Wharton}, {Wielgus}, {Wong}, {Wu}, {Young}, {Young}, {Younsi}, {Yuan}, {Yuan}, {Zensus}, {Zhao}, {Zhao}, {Zhu}, {Algaba}, {Allardi}, {Amestica}, {Anczarski}, {Bach}, {Baganoff}, {Beaudoin}, {Benson}, {Berthold}, {Blanchard}, {Blundell}, {Bustamente}, {Cappallo}, {Castillo-Dom{\'\i}nguez}, {Chang}, {Chang}, {Chang}, {Chen}, {Chilson}, {Chuter}, {C{\'o}rdova Rosado}, {Coulson}, {Crawford}, {Crowley}, {David}, {Derome}, {Dexter}, {Dornbusch}, {Dudevoir}, {Dzib}, {Eckart}, {Eckert}, {Erickson}, {Everett}, {Faber}, {Farah}, {Fath}, {Folkers}, {Forbes}, {Freund}, {G{\'o}mez-Ruiz}, {Gale}, {Gao}, {Geertsema}, {Graham}, {Greer}, {Grosslein}, {Gueth}, {Haggard}, {Halverson}, {Han}, {Han}, {Hao}, {Hasegawa}, {Henning}, {Hern{\'a}ndez-G{\'o}mez}, {Herrero-Illana}, {Heyminck}, {Hirota}, {Hoge}, {Huang}, {Impellizzeri}, {Jiang}, {Kamble}, {Keisler}, {Kimura}, {Kono}, {Kubo}, {Kuroda}, {Lacasse},
  {Laing}, {Leitch}, {Li}, {Lin}, {Liu}, {Liu}, {Lu}, {Marson}, {Martin-Cocher}, {Massingill}, {Matulonis}, {McColl}, {McWhirter}, {Messias}, {Meyer-Zhao}, {Michalik}, {Monta{\~n}a}, {Montgomerie}, {Mora-Klein}, {Muders}, {Nadolski}, {Navarro}, {Neilsen}, {Nguyen}, {Nishioka}, {Norton}, {Nowak}, {Nystrom}, {Ogawa}, {Oshiro}, {Oyama}, {Parsons}, {Paine}, {Pe{\~n}alver}, {Phillips}, {Poirier}, {Pradel}, {Primiani}, {Raffin}, {Rahlin}, {Reiland}, {Risacher}, {Ruiz}, {S{\'a}ez-Mada{\'\i}n}, {Sassella}, {Schellart}, {Shaw}, {Silva}, {Shiokawa}, {Smith}, {Snow}, {Souccar}, {Sousa}, {Sridharan}, {Srinivasan}, {Stahm}, {Stark}, {Story}, {Timmer}, {Vertatschitsch}, {Walther}, {Wei}, {Whitehorn}, {Whitney}, {Woody}, {Wouterloot}, {Wright}, {Yamaguchi}, {Yu}, {Zeballos}, {Zhang}, \& {Ziurys}}]{EHT_M87_I}
{Event Horizon Telescope Collaboration}, {Akiyama}, K., {Alberdi}, A., {et~al.} 2019{\natexlab{a}}, \apjl, 875, L1

\bibitem[{{Event Horizon Telescope Collaboration} {et~al.}(2019{\natexlab{b}}){Event Horizon Telescope Collaboration}, {Akiyama}, {Alberdi}, {Alef}, {Asada}, {Azulay}, {Baczko}, {Ball}, {Balokovi{\'c}}, {Barrett}, {Bintley}, {Blackburn}, {Boland}, {Bouman}, {Bower}, {Bremer}, {Brinkerink}, {Brissenden}, {Britzen}, {Broderick}, {Broguiere}, {Bronzwaer}, {Byun}, {Carlstrom}, {Chael}, {Chan}, {Chatterjee}, {Chatterjee}, {Chen}, {Chen}, {Cho}, {Christian}, {Conway}, {Cordes}, {Crew}, {Cui}, {Davelaar}, {De Laurentis}, {Deane}, {Dempsey}, {Desvignes}, {Dexter}, {Doeleman}, {Eatough}, {Falcke}, {Fish}, {Fomalont}, {Fraga-Encinas}, {Freeman}, {Friberg}, {Fromm}, {G{\'o}mez}, {Galison}, {Gammie}, {Garc{\'\i}a}, {Gentaz}, {Georgiev}, {Goddi}, {Gold}, {Gu}, {Gurwell}, {Hada}, {Hecht}, {Hesper}, {Ho}, {Ho}, {Honma}, {Huang}, {Huang}, {Hughes}, {Ikeda}, {Inoue}, {Issaoun}, {James}, {Jannuzi}, {Janssen}, {Jeter}, {Jiang}, {Johnson}, {Jorstad}, {Jung}, {Karami}, {Karuppusamy}, {Kawashima}, {Keating}, {Kettenis}, {Kim},
  {Kim}, {Kim}, {Kino}, {Koay}, {Koch}, {Koyama}, {Kramer}, {Kramer}, {Krichbaum}, {Kuo}, {Lauer}, {Lee}, {Li}, {Li}, {Lindqvist}, {Liu}, {Liuzzo}, {Lo}, {Lobanov}, {Loinard}, {Lonsdale}, {Lu}, {MacDonald}, {Mao}, {Markoff}, {Marrone}, {Marscher}, {Mart{\'\i}-Vidal}, {Matsushita}, {Matthews}, {Medeiros}, {Menten}, {Mizuno}, {Mizuno}, {Moran}, {Moriyama}, {Moscibrodzka}, {M{\"u}ller}, {Nagai}, {Nagar}, {Nakamura}, {Narayan}, {Narayanan}, {Natarajan}, {Neri}, {Ni}, {Noutsos}, {Okino}, {Olivares}, {Oyama}, {{\"O}zel}, {Palumbo}, {Patel}, {Pen}, {Pesce}, {Pi{\'e}tu}, {Plambeck}, {PopStefanija}, {Porth}, {Prather}, {Preciado-L{\'o}pez}, {Psaltis}, {Pu}, {Ramakrishnan}, {Rao}, {Rawlings}, {Raymond}, {Rezzolla}, {Ripperda}, {Roelofs}, {Rogers}, {Ros}, {Rose}, {Roshanineshat}, {Rottmann}, {Roy}, {Ruszczyk}, {Ryan}, {Rygl}, {S{\'a}nchez}, {S{\'a}nchez-Arguelles}, {Sasada}, {Savolainen}, {Schloerb}, {Schuster}, {Shao}, {Shen}, {Small}, {Sohn}, {SooHoo}, {Tazaki}, {Tiede}, {Tilanus}, {Titus}, {Toma}, {Torne}, {Trent},
  {Trippe}, {Tsuda}, {van Bemmel}, {van Langevelde}, {van Rossum}, {Wagner}, {Wardle}, {Weintroub}, {Wex}, {Wharton}, {Wielgus}, {Wong}, {Wu}, {Young}, {Young}, \& {Younsi}}]{EHT_M87_IV}
{Event Horizon Telescope Collaboration}, {Akiyama}, K., {Alberdi}, A., {et~al.} 2019{\natexlab{b}}, \apjl, 875, L4

\bibitem[{{Event Horizon Telescope Collaboration} {et~al.}(2019{\natexlab{c}}){Event Horizon Telescope Collaboration}, {Akiyama}, {Alberdi}, {Alef}, {Asada}, {Azulay}, {Baczko}, {Ball}, {Balokovi{\'c}}, {Barrett}, {Bintley}, {Blackburn}, {Boland}, {Bouman}, {Bower}, {Bremer}, {Brinkerink}, {Brissenden}, {Britzen}, {Broderick}, {Broguiere}, {Bronzwaer}, {Byun}, {Carlstrom}, {Chael}, {Chan}, {Chatterjee}, {Chatterjee}, {Chen}, {Chen}, {Cho}, {Christian}, {Conway}, {Cordes}, {Crew}, {Cui}, {Davelaar}, {De Laurentis}, {Deane}, {Dempsey}, {Desvignes}, {Dexter}, {Doeleman}, {Eatough}, {Falcke}, {Fish}, {Fomalont}, {Fraga-Encinas}, {Friberg}, {Fromm}, {G{\'o}mez}, {Galison}, {Gammie}, {Garc{\'\i}a}, {Gentaz}, {Georgiev}, {Goddi}, {Gold}, {Gu}, {Gurwell}, {Hada}, {Hecht}, {Hesper}, {Ho}, {Ho}, {Honma}, {Huang}, {Huang}, {Hughes}, {Ikeda}, {Inoue}, {Issaoun}, {James}, {Jannuzi}, {Janssen}, {Jeter}, {Jiang}, {Johnson}, {Jorstad}, {Jung}, {Karami}, {Karuppusamy}, {Kawashima}, {Keating}, {Kettenis}, {Kim}, {Kim}, {Kim},
  {Kino}, {Koay}, {Koch}, {Koyama}, {Kramer}, {Kramer}, {Krichbaum}, {Kuo}, {Lauer}, {Lee}, {Li}, {Li}, {Lindqvist}, {Liu}, {Liuzzo}, {Lo}, {Lobanov}, {Loinard}, {Lonsdale}, {Lu}, {MacDonald}, {Mao}, {Markoff}, {Marrone}, {Marscher}, {Mart{\'\i}-Vidal}, {Matsushita}, {Matthews}, {Medeiros}, {Menten}, {Mizuno}, {Mizuno}, {Moran}, {Moriyama}, {Moscibrodzka}, {M{\"u}ller}, {Nagai}, {Nagar}, {Nakamura}, {Narayan}, {Narayanan}, {Natarajan}, {Neri}, {Ni}, {Noutsos}, {Okino}, {Olivares}, {Oyama}, {{\"O}zel}, {Palumbo}, {Patel}, {Pen}, {Pesce}, {Pi{\'e}tu}, {Plambeck}, {PopStefanija}, {Porth}, {Prather}, {Preciado-L{\'o}pez}, {Psaltis}, {Pu}, {Ramakrishnan}, {Rao}, {Rawlings}, {Raymond}, {Rezzolla}, {Ripperda}, {Roelofs}, {Rogers}, {Ros}, {Rose}, {Roshanineshat}, {Rottmann}, {Roy}, {Ruszczyk}, {Ryan}, {Rygl}, {S{\'a}nchez}, {S{\'a}nchez-Arguelles}, {Sasada}, {Savolainen}, {Schloerb}, {Schuster}, {Shao}, {Shen}, {Small}, {Sohn}, {SooHoo}, {Tazaki}, {Tiede}, {Tilanus}, {Titus}, {Toma}, {Torne}, {Trent}, {Trippe},
  {Tsuda}, {van Bemmel}, {van Langevelde}, {van Rossum}, {Wagner}, {Wardle}, {Weintroub}, {Wex}, {Wharton}, {Wielgus}, {Wong}, {Wu}, {Young}, {Young}, {Younsi}, {Yuan}, {Yuan}, {Zensus}, {Zhao}, {Zhao}, {Zhu}, {Farah}, {Meyer-Zhao}, {Michalik}, {Nadolski}, {Nishioka}, {Pradel}, {Primiani}, {Souccar}, {Vertatschitsch}, \& {Yamaguchi}}]{EHT_M87_VI}
{Event Horizon Telescope Collaboration}, {Akiyama}, K., {Alberdi}, A., {et~al.} 2019{\natexlab{c}}, \apjl, 875, L6

\bibitem[{{Event Horizon Telescope Collaboration} {et~al.}(2019{\natexlab{d}}){Event Horizon Telescope Collaboration}, {Akiyama}, {Alberdi}, {Alef}, {Asada}, {Azulay}, {Baczko}, {Ball}, {Balokovi{\'c}}, {Barrett}, {Bintley}, {Blackburn}, {Boland}, {Bouman}, {Bower}, {Bremer}, {Brinkerink}, {Brissenden}, {Britzen}, {Broderick}, {Broguiere}, {Bronzwaer}, {Byun}, {Carlstrom}, {Chael}, {Chan}, {Chatterjee}, {Chatterjee}, {Chen}, {Chen}, {Cho}, {Christian}, {Conway}, {Cordes}, {Crew}, {Cui}, {Davelaar}, {De Laurentis}, {Deane}, {Dempsey}, {Desvignes}, {Dexter}, {Doeleman}, {Eatough}, {Falcke}, {Fish}, {Fomalont}, {Fraga-Encinas}, {Friberg}, {Fromm}, {G{\'o}mez}, {Galison}, {Gammie}, {Garc{\'\i}a}, {Gentaz}, {Georgiev}, {Goddi}, {Gold}, {Gu}, {Gurwell}, {Hada}, {Hecht}, {Hesper}, {Ho}, {Ho}, {Honma}, {Huang}, {Huang}, {Hughes}, {Ikeda}, {Inoue}, {Issaoun}, {James}, {Jannuzi}, {Janssen}, {Jeter}, {Jiang}, {Johnson}, {Jorstad}, {Jung}, {Karami}, {Karuppusamy}, {Kawashima}, {Keating}, {Kettenis}, {Kim}, {Kim}, {Kim},
  {Kino}, {Koay}, {Koch}, {Koyama}, {Kramer}, {Kramer}, {Krichbaum}, {Kuo}, {Lauer}, {Lee}, {Li}, {Li}, {Lindqvist}, {Liu}, {Liuzzo}, {Lo}, {Lobanov}, {Loinard}, {Lonsdale}, {Lu}, {MacDonald}, {Mao}, {Markoff}, {Marrone}, {Marscher}, {Mart{\'\i}-Vidal}, {Matsushita}, {Matthews}, {Medeiros}, {Menten}, {Mizuno}, {Mizuno}, {Moran}, {Moriyama}, {Moscibrodzka}, {M{\"u}ller}, {Nagai}, {Nagar}, {Nakamura}, {Narayan}, {Narayanan}, {Natarajan}, {Neri}, {Ni}, {Noutsos}, {Okino}, {Olivares}, {Ortiz-Le{\'o}n}, {Oyama}, {{\"O}zel}, {Palumbo}, {Patel}, {Pen}, {Pesce}, {Pi{\'e}tu}, {Plambeck}, {PopStefanija}, {Porth}, {Prather}, {Preciado-L{\'o}pez}, {Psaltis}, {Pu}, {Ramakrishnan}, {Rao}, {Rawlings}, {Raymond}, {Rezzolla}, {Ripperda}, {Roelofs}, {Rogers}, {Ros}, {Rose}, {Roshanineshat}, {Rottmann}, {Roy}, {Ruszczyk}, {Ryan}, {Rygl}, {S{\'a}nchez}, {S{\'a}nchez-Arguelles}, {Sasada}, {Savolainen}, {Schloerb}, {Schuster}, {Shao}, {Shen}, {Small}, {Sohn}, {SooHoo}, {Tazaki}, {Tiede}, {Tilanus}, {Titus}, {Toma}, {Torne},
  {Trent}, {Trippe}, {Tsuda}, {van Bemmel}, {van Langevelde}, {van Rossum}, {Wagner}, {Wardle}, {Weintroub}, {Wex}, {Wharton}, {Wielgus}, {Wong}, {Wu}, {Young}, {Young}, \& {Younsi}}]{EHT_M87_II}
{Event Horizon Telescope Collaboration}, {Akiyama}, K., {Alberdi}, A., {et~al.} 2019{\natexlab{d}}, \apjl, 875, L2

\bibitem[{{Event Horizon Telescope Collaboration} {et~al.}(2021{\natexlab{a}}){Event Horizon Telescope Collaboration}, {Akiyama}, {Algaba}, {Alberdi}, {Alef}, {Anantua}, {Asada}, {Azulay}, {Baczko}, {Ball}, {Balokovi{\'c}}, {Barrett}, {Benson}, {Bintley}, {Blackburn}, {Blundell}, {Boland}, {Bouman}, {Bower}, {Boyce}, {Bremer}, {Brinkerink}, {Brissenden}, {Britzen}, {Broderick}, {Broguiere}, {Bronzwaer}, {Byun}, {Carlstrom}, {Chael}, {Chan}, {Chatterjee}, {Chatterjee}, {Chen}, {Chen}, {Chesler}, {Cho}, {Christian}, {Conway}, {Cordes}, {Crawford}, {Crew}, {Cruz-Osorio}, {Cui}, {Davelaar}, {De Laurentis}, {Deane}, {Dempsey}, {Desvignes}, {Dexter}, {Doeleman}, {Eatough}, {Falcke}, {Farah}, {Fish}, {Fomalont}, {Ford}, {Fraga-Encinas}, {Freeman}, {Friberg}, {Fromm}, {Fuentes}, {Galison}, {Gammie}, {Garc{\'\i}a}, {Gentaz}, {Georgiev}, {Goddi}, {Gold}, {G{\'o}mez}, {G{\'o}mez-Ruiz}, {Gu}, {Gurwell}, {Hada}, {Haggard}, {Hecht}, {Hesper}, {Ho}, {Ho}, {Honma}, {Huang}, {Huang}, {Hughes}, {Ikeda}, {Inoue}, {Issaoun},
  {James}, {Jannuzi}, {Janssen}, {Jeter}, {Jiang}, {Jimenez-Rosales}, {Johnson}, {Jorstad}, {Jung}, {Karami}, {Karuppusamy}, {Kawashima}, {Keating}, {Kettenis}, {Kim}, {Kim}, {Kim}, {Kim}, {Kino}, {Koay}, {Kofuji}, {Koch}, {Koyama}, {Kramer}, {Kramer}, {Krichbaum}, {Kuo}, {Lauer}, {Lee}, {Levis}, {Li}, {Li}, {Lindqvist}, {Lico}, {Lindahl}, {Liu}, {Liu}, {Liuzzo}, {Lo}, {Lobanov}, {Loinard}, {Lonsdale}, {Lu}, {MacDonald}, {Mao}, {Marchili}, {Markoff}, {Marrone}, {Marscher}, {Mart{\'\i}-Vidal}, {Matsushita}, {Matthews}, {Medeiros}, {Menten}, {Mizuno}, {Mizuno}, {Moran}, {Moriyama}, {Moscibrodzka}, {M{\"u}ller}, {Musoke}, {Mej{\'\i}as}, {Michalik}, {Nadolski}, {Nagai}, {Nagar}, {Nakamura}, {Narayan}, {Narayanan}, {Natarajan}, {Nathanail}, {Neilsen}, {Neri}, {Ni}, {Noutsos}, {Nowak}, {Okino}, {Olivares}, {Ortiz-Le{\'o}n}, {Oyama}, {{\"O}zel}, {Palumbo}, {Park}, {Patel}, {Pen}, {Pesce}, {Pi{\'e}tu}, {Plambeck}, {PopStefanija}, {Porth}, {P{\"o}tzl}, {Prather}, {Preciado-L{\'o}pez}, {Psaltis}, {Pu}, {Ramakrishnan},
  {Rao}, {Rawlings}, {Raymond}, {Rezzolla}, {Ricarte}, {Ripperda}, {Roelofs}, {Rogers}, {Ros}, {Rose}, {Roshanineshat}, {Rottmann}, {Roy}, {Ruszczyk}, {Rygl}, {S{\'a}nchez}, {S{\'a}nchez-Arguelles}, {Sasada}, {Savolainen}, {Schloerb}, {Schuster}, {Shao}, {Shen}, {Small}, {Sohn}, {SooHoo}, {Sun}, {Tazaki}, {Tetarenko}, {Tiede}, {Tilanus}, {Titus}, {Toma}, {Torne}, {Trent}, {Traianou}, {Trippe}, {van Bemmel}, {van Langevelde}, {van Rossum}, {Wagner}, {Ward-Thompson}, {Wardle}, {Weintroub}, {Wex}, {Wharton}, {Wielgus}, {Wong}, {Wu}, {Yoon}, {Young}, {Young}, {Younsi}, {Yuan}, {Yuan}, {Zensus}, {Zhao}, \& {Zhao}}]{EHT_M87_VII}
{Event Horizon Telescope Collaboration}, {Akiyama}, K., {Algaba}, J.~C., {et~al.} 2021{\natexlab{a}}, \apjl, 910, L12

\bibitem[{{Event Horizon Telescope Collaboration} {et~al.}(2021{\natexlab{b}}){Event Horizon Telescope Collaboration}, {Akiyama}, {Algaba}, {Alberdi}, {Alef}, {Anantua}, {Asada}, {Azulay}, {Baczko}, {Ball}, {Balokovi{\'c}}, {Barrett}, {Benson}, {Bintley}, {Blackburn}, {Blundell}, {Boland}, {Bouman}, {Bower}, {Boyce}, {Bremer}, {Brinkerink}, {Brissenden}, {Britzen}, {Broderick}, {Broguiere}, {Bronzwaer}, {Byun}, {Carlstrom}, {Chael}, {Chan}, {Chatterjee}, {Chatterjee}, {Chen}, {Chen}, {Chesler}, {Cho}, {Christian}, {Conway}, {Cordes}, {Crawford}, {Crew}, {Cruz-Osorio}, {Cui}, {Davelaar}, {De Laurentis}, {Deane}, {Dempsey}, {Desvignes}, {Dexter}, {Doeleman}, {Eatough}, {Falcke}, {Farah}, {Fish}, {Fomalont}, {Ford}, {Fraga-Encinas}, {Friberg}, {Fromm}, {Fuentes}, {Galison}, {Gammie}, {Garc{\'\i}a}, {Gelles}, {Gentaz}, {Georgiev}, {Goddi}, {Gold}, {G{\'o}mez}, {G{\'o}mez-Ruiz}, {Gu}, {Gurwell}, {Hada}, {Haggard}, {Hecht}, {Hesper}, {Himwich}, {Ho}, {Ho}, {Honma}, {Huang}, {Huang}, {Hughes}, {Ikeda}, {Inoue},
  {Issaoun}, {James}, {Jannuzi}, {Janssen}, {Jeter}, {Jiang}, {Jimenez-Rosales}, {Johnson}, {Jorstad}, {Jung}, {Karami}, {Karuppusamy}, {Kawashima}, {Keating}, {Kettenis}, {Kim}, {Kim}, {Kim}, {Kim}, {Kino}, {Koay}, {Kofuji}, {Koch}, {Koyama}, {Kramer}, {Kramer}, {Krichbaum}, {Kuo}, {Lauer}, {Lee}, {Levis}, {Li}, {Li}, {Lindqvist}, {Lico}, {Lindahl}, {Liu}, {Liu}, {Liuzzo}, {Lo}, {Lobanov}, {Loinard}, {Lonsdale}, {Lu}, {MacDonald}, {Mao}, {Marchili}, {Markoff}, {Marrone}, {Marscher}, {Mart{\'\i}-Vidal}, {Matsushita}, {Matthews}, {Medeiros}, {Menten}, {Mizuno}, {Mizuno}, {Moran}, {Moriyama}, {Moscibrodzka}, {M{\"u}ller}, {Musoke}, {Mus Mej{\'\i}as}, {Michalik}, {Nadolski}, {Nagai}, {Nagar}, {Nakamura}, {Narayan}, {Narayanan}, {Natarajan}, {Nathanail}, {Neilsen}, {Neri}, {Ni}, {Noutsos}, {Nowak}, {Okino}, {Olivares}, {Ortiz-Le{\'o}n}, {Oyama}, {{\"O}zel}, {Palumbo}, {Park}, {Patel}, {Pen}, {Pesce}, {Pi{\'e}tu}, {Plambeck}, {PopStefanija}, {Porth}, {P{\"o}tzl}, {Prather}, {Preciado-L{\'o}pez}, {Psaltis}, {Pu},
  {Ramakrishnan}, {Rao}, {Rawlings}, {Raymond}, {Rezzolla}, {Ricarte}, {Ripperda}, {Roelofs}, {Rogers}, {Ros}, {Rose}, {Roshanineshat}, {Rottmann}, {Roy}, {Ruszczyk}, {Rygl}, {S{\'a}nchez}, {S{\'a}nchez-Arguelles}, {Sasada}, {Savolainen}, {Schloerb}, {Schuster}, {Shao}, {Shen}, {Small}, {Sohn}, {SooHoo}, {Sun}, {Tazaki}, {Tetarenko}, {Tiede}, {Tilanus}, {Titus}, {Toma}, {Torne}, {Trent}, {Traianou}, {Trippe}, {van Bemmel}, {van Langevelde}, {van Rossum}, {Wagner}, {Ward-Thompson}, {Wardle}, {Weintroub}, {Wex}, {Wharton}, {Wielgus}, {Wong}, {Wu}, {Yoon}, {Young}, {Young}, {Younsi}, {Yuan}, {Yuan}, {Zensus}, {Zhao}, \& {Zhao}}]{EHT_M87_VIII}
{Event Horizon Telescope Collaboration}, {Akiyama}, K., {Algaba}, J.~C., {et~al.} 2021{\natexlab{b}}, \apjl, 910, L13

\bibitem[{{Filothodoros} {et~al.}(2024){Filothodoros}, {Lewandowski}, {Kijak}, {{\'S}mierciak}, {Chy{\.z}y}, {B{\l}aszkiewicz}, \& {Krankowski}}]{Filoth:2024}
{Filothodoros}, A., {Lewandowski}, W., {Kijak}, J., {et~al.} 2024, \mnras, 528, 5667

\bibitem[{{Fish} {et~al.}(2020){Fish}, {Shea}, \& {Akiyama}}]{Fish:2020}
{Fish}, V.~L., {Shea}, M., \& {Akiyama}, K. 2020, Advances in Space Research, 65, 821

\bibitem[{{Fromm} {et~al.}(2021){Fromm}, {Mizuno}, {Younsi}, {Olivares}, {Porth}, {De Laurentis}, {Falcke}, {Kramer}, \& {Rezzolla}}]{Fromm:2021}
{Fromm}, C.~M., {Mizuno}, Y., {Younsi}, Z., {et~al.} 2021, \aap, 649, A116

\bibitem[{{Gab{\'a}nyi} {et~al.}(2006){Gab{\'a}nyi}, {Krichbaum}, {Britzen}, {Bach}, {Ros}, {Witzel}, \& {Zensus}}]{Gabanyi:2006}
{Gab{\'a}nyi}, K.~{\'E}., {Krichbaum}, T.~P., {Britzen}, S., {et~al.} 2006, \aap, 451, 85

\bibitem[{{Ghez} {et~al.}(2008){Ghez}, {Salim}, {Weinberg}, {Lu}, {Do}, {Dunn}, {Matthews}, {Morris}, {Yelda}, {Becklin}, {Kremenek}, {Milosavljevic}, \& {Naiman}}]{Ghez:2008}
{Ghez}, A.~M., {Salim}, S., {Weinberg}, N.~N., {et~al.} 2008, \apj, 689, 1044

\bibitem[{{Goddi} {et~al.}(2021){Goddi}, {Mart{\'\i}-Vidal}, {Messias}, {Bower}, {Broderick}, {Dexter}, {Marrone}, {Moscibrodzka}, {Nagai}, {Algaba}, {Asada}, {Crew}, {G{\'o}mez}, {Impellizzeri}, {Janssen}, {Kadler}, {Krichbaum}, {Lico}, {Matthews}, {Nathanail}, {Ricarte}, {Ros}, {Younsi}, {Akiyama}, {Alberdi}, {Alef}, {Anantua}, {Azulay}, {Baczko}, {Ball}, {Balokovi{\'c}}, {Barrett}, {Benson}, {Bintley}, {Blackburn}, {Blundell}, {Boland}, {Bouman}, {Boyce}, {Bremer}, {Brinkerink}, {Brissenden}, {Britzen}, {Broguiere}, {Bronzwaer}, {Byun}, {Carlstrom}, {Chael}, {Chan}, {Chatterjee}, {Chatterjee}, {Chen}, {Chen}, {Chesler}, {Cho}, {Christian}, {Conway}, {Cordes}, {Crawford}, {Cruz-Osorio}, {Cui}, {Davelaar}, {De Laurentis}, {Deane}, {Dempsey}, {Desvignes}, {Doeleman}, {Eatough}, {Falcke}, {Farah}, {Fish}, {Fomalont}, {Ford}, {Fraga-Encinas}, {Freeman}, {Friberg}, {Fromm}, {Fuentes}, {Galison}, {Gammie}, {Garc{\'\i}a}, {Gentaz}, {Georgiev}, {Gold}, {G{\'o}mez-Ruiz}, {Gu}, {Gurwell}, {Hada}, {Haggard}, {Hecht},
  {Hesper}, {Ho}, {Ho}, {Honma}, {Huang}, {Huang}, {Hughes}, {Inoue}, {Issaoun}, {James}, {Jannuzi}, {Jeter}, {Jiang}, {Jimenez-Rosales}, {Johnson}, {Jorstad}, {Jung}, {Karami}, {Karuppusamy}, {Kawashima}, {Keating}, {Kettenis}, {Kim}, {Kim}, {Kim}, {Kim}, {Kino}, {Koay}, {Kofuji}, {Koch}, {Koyama}, {Kramer}, {Kramer}, {Kuo}, {Lauer}, {Lee}, {Levis}, {Li}, {Li}, {Lindqvist}, {Lindahl}, {Liu}, {Liu}, {Liuzzo}, {Lo}, {Lobanov}, {Loinard}, {Lonsdale}, {Lu}, {MacDonald}, {Mao}, {Marchili}, {Markoff}, {Marscher}, {Matsushita}, {Medeiros}, {Menten}, {Mizuno}, {Mizuno}, {Moran}, {Moriyama}, {M{\"u}ller}, {Musoke}, {Mej{\'\i}as}, {Nagar}, {Nakamura}, {Narayan}, {Narayanan}, {Natarajan}, {Neilsen}, {Neri}, {Ni}, {Noutsos}, {Nowak}, {Okino}, {Olivares}, {Ortiz-Le{\'o}n}, {Oyama}, {{\"O}zel}, {Palumbo}, {Park}, {Patel}, {Pen}, {Pesce}, {Pi{\'e}tu}, {Plambeck}, {PopStefanija}, {Porth}, {P{\"o}tzl}, {Prather}, {Preciado-L{\'o}pez}, {Psaltis}, {Pu}, {Ramakrishnan}, {Rao}, {Rawlings}, {Raymond}, {Rezzolla}, {Ripperda},
  {Roelofs}, {Rogers}, {Rose}, {Roshanineshat}, {Rottmann}, {Roy}, {Ruszczyk}, {Rygl}, {S{\'a}nchez}, {S{\'a}nchez-Arguelles}, {Sasada}, {Savolainen}, {Schloerb}, {Schuster}, {Shao}, {Shen}, {Small}, {Sohn}, {SooHoo}, {Sun}, {Tazaki}, {Tetarenko}, {Tiede}, {Tilanus}, {Titus}, {Toma}, {Torne}, {Trent}, {Traianou}, {Trippe}, {van Bemmel}, {van Langevelde}, {van Rossum}, {Wagner}, {Ward-Thompson}, {Wardle}, {Weintroub}, {Wex}, {Wharton}, {Wielgus}, {Wong}, {Wu}, {Yoon}, {Young}, {Young}, {Yuan}, {Yuan}, {Zensus}, {Zhao}, {Zhao}, {Bruni}, {Gopakumar}, {Hern{\'a}ndez-G{\'o}mez}, {Herrero-Illana}, {Ingram}, {Komossa}, {Kovalev}, {Muders}, {Perucho}, {R{\"o}sch}, \& {Valtonen}}]{Goddi:2021}
{Goddi}, C., {Mart{\'\i}-Vidal}, I., {Messias}, H., {et~al.} 2021, \apjl, 910, L14

\bibitem[{{Goldreich} \& {Sridhar}(1995)}]{Goldrecih:1995}
{Goldreich}, P. \& {Sridhar}, S. 1995, \apj, 438, 763

\bibitem[{{Goodman} \& {Narayan}(1989)}]{Goodman:1989}
{Goodman}, J. \& {Narayan}, R. 1989, \mnras, 238, 995

\bibitem[{{Gralla} \& {Lupsasca}(2020)}]{Gralla:2020A}
{Gralla}, S.~E. \& {Lupsasca}, A. 2020, \prd, 102, 124003

\bibitem[{{Gralla} {et~al.}(2020{\natexlab{a}}){Gralla}, {Lupsasca}, \& {Marrone}}]{Gralla:2020}
{Gralla}, S.~E., {Lupsasca}, A., \& {Marrone}, D.~P. 2020{\natexlab{a}}, \prd, 102, 124004

\bibitem[{{Gralla} {et~al.}(2020{\natexlab{b}}){Gralla}, {Lupsasca}, \& {Marrone}}]{GLM_2020}
{Gralla}, S.~E., {Lupsasca}, A., \& {Marrone}, D.~P. 2020{\natexlab{b}}, \prd, 102, 124004

\bibitem[{{GRAVITY Collaboration} {et~al.}(2018){GRAVITY Collaboration}, {Abuter}, {Amorim}, {Anugu}, {Baub{\"o}ck}, {Benisty}, {Berger}, {Blind}, {Bonnet}, {Brandner}, {Buron}, {Collin}, {Chapron}, {Cl{\'e}net}, {Coud{\'e} Du Foresto}, {de Zeeuw}, {Deen}, {Delplancke-Str{\"o}bele}, {Dembet}, {Dexter}, {Duvert}, {Eckart}, {Eisenhauer}, {Finger}, {F{\"o}rster Schreiber}, {F{\'e}dou}, {Garcia}, {Garcia Lopez}, {Gao}, {Gendron}, {Genzel}, {Gillessen}, {Gordo}, {Habibi}, {Haubois}, {Haug}, {Hau{\ss}mann}, {Henning}, {Hippler}, {Horrobin}, {Hubert}, {Hubin}, {Jimenez Rosales}, {Jochum}, {Jocou}, {Kaufer}, {Kellner}, {Kendrew}, {Kervella}, {Kok}, {Kulas}, {Lacour}, {Lapeyr{\`e}re}, {Lazareff}, {Le Bouquin}, {L{\'e}na}, {Lippa}, {Lenzen}, {M{\'e}rand}, {M{\"u}ler}, {Neumann}, {Ott}, {Palanca}, {Paumard}, {Pasquini}, {Perraut}, {Perrin}, {Pfuhl}, {Plewa}, {Rabien}, {Ram{\'\i}rez}, {Ramos}, {Rau}, {Rodr{\'\i}guez-Coira}, {Rohloff}, {Rousset}, {Sanchez-Bermudez}, {Scheithauer}, {Sch{\"o}ller}, {Schuler}, {Spyromilio},
  {Straub}, {Straubmeier}, {Sturm}, {Tacconi}, {Tristram}, {Vincent}, {von Fellenberg}, {Wank}, {Waisberg}, {Widmann}, {Wieprecht}, {Wiest}, {Wiezorrek}, {Woillez}, {Yazici}, {Ziegler}, \& {Zins}}]{GRAVITY:2018}
{GRAVITY Collaboration}, {Abuter}, R., {Amorim}, A., {et~al.} 2018, \aap, 615, L15

\bibitem[{{GRAVITY Collaboration} {et~al.}(2019){GRAVITY Collaboration}, {Abuter}, {Amorim}, {Baub{\"o}ck}, {Berger}, {Bonnet}, {Brandner}, {Cl{\'e}net}, {Coud{\'e} Du Foresto}, {de Zeeuw}, {Dexter}, {Duvert}, {Eckart}, {Eisenhauer}, {F{\"o}rster Schreiber}, {Garcia}, {Gao}, {Gendron}, {Genzel}, {Gerhard}, {Gillessen}, {Habibi}, {Haubois}, {Henning}, {Hippler}, {Horrobin}, {Jim{\'e}nez-Rosales}, {Jocou}, {Kervella}, {Lacour}, {Lapeyr{\`e}re}, {Le Bouquin}, {L{\'e}na}, {Ott}, {Paumard}, {Perraut}, {Perrin}, {Pfuhl}, {Rabien}, {Rodriguez Coira}, {Rousset}, {Scheithauer}, {Sternberg}, {Straub}, {Straubmeier}, {Sturm}, {Tacconi}, {Vincent}, {von Fellenberg}, {Waisberg}, {Widmann}, {Wieprecht}, {Wiezorrek}, {Woillez}, \& {Yazici}}]{GRAVITY:2019}
{GRAVITY Collaboration}, {Abuter}, R., {Amorim}, A., {et~al.} 2019, \aap, 625, L10

\bibitem[{{Gupta} {et~al.}(1993){Gupta}, {Rickett}, \& {Coles}}]{Gupta:1993}
{Gupta}, Y., {Rickett}, B.~J., \& {Coles}, W.~A. 1993, \apj, 403, 183

\bibitem[{{Gurvits}(2020)}]{Gurvits:2020}
{Gurvits}, L.~I. 2020, Advances in Space Research, 65, 868

\bibitem[{{Gwinn} {et~al.}(2014){Gwinn}, {Kovalev}, {Johnson}, \& {Soglasnov}}]{Gwinn:2014}
{Gwinn}, C.~R., {Kovalev}, Y.~Y., {Johnson}, M.~D., \& {Soglasnov}, V.~A. 2014, \apjl, 794, L14

\bibitem[{Hart {et~al.}(2018)Hart, Brown, Collins, De~Soria-Santacruz~Pich, Fieseler, Goebel, Marsh, Oh, Snyder, Warner, Whiffen, Elkins-Tanton, Bell, Lawrence, Lord, \& Pirkl}]{Hart:2018}
Hart, W., Brown, G.~M., Collins, S.~M., {et~al.} 2018, in 2018 IEEE Aerospace Conference, 1--20

\bibitem[{Hudson(2024)}]{hudson_python_2024}
Hudson, B. 2024, Python package for simulating and optimising a space-based {VLBI} mission, language: en Medium: Python package

\bibitem[{{Hudson} {et~al.}(2023){Hudson}, Gurvits, Wielgus, Paragi, Liu, \& Zheng}]{Hudson_2023}
{Hudson}, B., Gurvits, L.~I., Wielgus, M., {et~al.} 2023, Acta Astronautica, 213, 681

\bibitem[{{Issaoun} {et~al.}(2019){Issaoun}, {Johnson}, {Blackburn}, {Brinkerink}, {Mo{\'s}cibrodzka}, {Chael}, {Goddi}, {Mart{\'\i}-Vidal}, {Wagner}, {Doeleman}, {Falcke}, {Krichbaum}, {Akiyama}, {Bach}, {Bouman}, {Bower}, {Broderick}, {Cho}, {Crew}, {Dexter}, {Fish}, {Gold}, {G{\'o}mez}, {Hada}, {Hern{\'a}ndez-G{\'o}mez}, {Jan{\ss}en}, {Kino}, {Kramer}, {Loinard}, {Lu}, {Markoff}, {Marrone}, {Matthews}, {Moran}, {M{\"u}ller}, {Roelofs}, {Ros}, {Rottmann}, {Sanchez}, {Tilanus}, {de Vicente}, {Wielgus}, {Zensus}, \& {Zhao}}]{Issaoun:2019}
{Issaoun}, S., {Johnson}, M.~D., {Blackburn}, L., {et~al.} 2019, \apj, 871, 30

\bibitem[{{Jensen} {et~al.}(2003){Jensen}, {Clausen}, {Cassi}, {Ravera}, {Janin}, {Winkler}, \& {Much}}]{Jensen:2003}
{Jensen}, P.~L., {Clausen}, K., {Cassi}, C., {et~al.} 2003, \aap, 411, L7

\bibitem[{{Jim{\'e}nez-Rosales} {et~al.}(2021){Jim{\'e}nez-Rosales}, {Dexter}, {Ressler}, {Tchekhovskoy}, {Baub{\"o}ck}, {Dallilar}, {de Zeeuw}, {Drescher}, {Eisenhauer}, {von Fellenberg}, {Gao}, {Genzel}, {Gillessen}, {Habibi}, {Ott}, {Stadler}, {Straub}, \& {Widmann}}]{Rosales:2021}
{Jim{\'e}nez-Rosales}, A., {Dexter}, J., {Ressler}, S.~M., {et~al.} 2021, \mnras, 503, 4563

\bibitem[{{Johnson}(2016)}]{Johnson:2016A}
{Johnson}, M.~D. 2016, \apj, 833, 74

\bibitem[{{Johnson} {et~al.}(2024){Johnson}, {Akiyama}, {Baturin}, {Bilyeu}, {Blackburn}, {Boroson}, {Cardenas-Avendano}, {Chael}, {Chan}, {Chang}, {Cheimets}, {Chou}, {Doeleman}, {Farah}, {Galison}, {Gamble}, {Gammie}, {Gelles}, {Gomez}, {Gralla}, {Grimes}, {Gurvits}, {Hadar}, {Haworth}, {Hada}, {Hecht}, {Honma}, {Houston}, {Hudson}, {Issaoun}, {Jia}, {Jorstad}, {Kauffmann}, {Kovalev}, {Kurczynski}, {Lafon}, {Lupsasca}, {Lehmensiek}, {Ma}, {Marrone}, {Marscher}, {Melnick}, {Narayan}, {Niinuma}, {Noble}, {Palmer}, {Palumbo}, {Paritsky}, {Peretz}, {Pesce}, {Plavin}, {Quataert}, {Rana}, {Ricarte}, {Roelofs}, {Shtyrkova}, {Sinclair}, {Small}, {Tirupati Kumara}, {Srinivasan}, {Strominger}, {Tiede}, {Tong}, {Wang}, {Weintroub}, {Wielgus}, {Wong}, \& {Zhang}}]{Johnson:2024}
{Johnson}, M.~D., {Akiyama}, K., {Baturin}, R., {et~al.} 2024, arXiv e-prints, arXiv:2406.12917

\bibitem[{{Johnson} \& {Gwinn}(2015)}]{Johnson_GW:2015}
{Johnson}, M.~D. \& {Gwinn}, C.~R. 2015, \apj, 805, 180

\bibitem[{{Johnson} {et~al.}(2021){Johnson}, {Kovalev}, {Lisakov}, {Voitsik}, {Gwinn}, \& {Bruni}}]{Johnson:2021}
{Johnson}, M.~D., {Kovalev}, Y.~Y., {Lisakov}, M.~M., {et~al.} 2021, \apjl, 922, L28

\bibitem[{{Johnson} {et~al.}(2020){Johnson}, {Lupsasca}, {Strominger}, {Wong}, {Hadar}, {Kapec}, {Narayan}, {Chael}, {Gammie}, {Galison}, {Palumbo}, {Doeleman}, {Blackburn}, {Wielgus}, {Pesce}, {Farah}, \& {Moran}}]{Johnson:2020}
{Johnson}, M.~D., {Lupsasca}, A., {Strominger}, A., {et~al.} 2020, Science Advances, 6, eaaz1310

\bibitem[{{Johnson} \& {Narayan}(2016)}]{Johnson:2016}
{Johnson}, M.~D. \& {Narayan}, R. 2016, \apj, 826, 170

\bibitem[{{Johnson} {et~al.}(2018){Johnson}, {Narayan}, {Psaltis}, {Blackburn}, {Kovalev}, {Gwinn}, {Zhao}, {Bower}, {Moran}, {Kino}, {Kramer}, {Akiyama}, {Dexter}, {Broderick}, \& {Sironi}}]{Johnson:2018}
{Johnson}, M.~D., {Narayan}, R., {Psaltis}, D., {et~al.} 2018, \apj, 865, 104

\bibitem[{Khan {et~al.}(2024)Khan, Lawler, Nicholas, Ortiz, Mukherji, Nowicki, Redfield, Voskanian, Mak, Ruderman, \& Mallamaci}]{Khan:2024}
Khan, S., Lawler, C., Nicholas, A., {et~al.} 2024, in 2024 IEEE Aerospace Conference, 1--17

\bibitem[{{Koryukova} {et~al.}(2022){Koryukova}, {Pushkarev}, {Plavin}, \& {Kovalev}}]{Koryukova:2022}
{Koryukova}, T.~A., {Pushkarev}, A.~B., {Plavin}, A.~V., \& {Kovalev}, Y.~Y. 2022, \mnras, 515, 1736

\bibitem[{Lambert \& Rickett(2000)}]{Lambert:2000}
Lambert, H.~C. \& Rickett, B.~J. 2000, \apj, 531, 883

\bibitem[{{Likhachev} {et~al.}(2022){Likhachev}, {Rudnitskiy}, {Shchurov}, {Andrianov}, {Baryshev}, {Chernov}, \& {Kostenko}}]{Likhachev:2022}
{Likhachev}, S.~F., {Rudnitskiy}, A.~G., {Shchurov}, M.~A., {et~al.} 2022, \mnras, 511, 668

\bibitem[{{Lo} {et~al.}(1993){Lo}, {Backer}, {Kellermann}, {Reid}, {Zhao}, {Goss}, \& {Moran}}]{Lo:1993}
{Lo}, K.~Y., {Backer}, D.~C., {Kellermann}, K.~I., {et~al.} 1993, \nat, 362, 38

\bibitem[{Mailhe \& Heister(2002)}]{Mailhe:2022}
Mailhe, L.~M. \& Heister, S.~D. 2002, Journal of Spacecraft and Rockets, 39, 131

\bibitem[{{Narayan} \& {Goodman}(1989)}]{Narayan:1989}
{Narayan}, R. \& {Goodman}, J. 1989, \mnras, 238, 963

\bibitem[{{{\"O}zel} {et~al.}(2022){{\"O}zel}, {Psaltis}, \& {Younsi}}]{Ozel:2022}
{{\"O}zel}, F., {Psaltis}, D., \& {Younsi}, Z. 2022, \apj, 941, 88

\bibitem[{{Palumbo} {et~al.}(2019{\natexlab{a}}){Palumbo}, {Doeleman}, {Johnson}, {Bouman}, \& {Chael}}]{Palumbo:2019}
{Palumbo}, D. C.~M., {Doeleman}, S.~S., {Johnson}, M.~D., {Bouman}, K.~L., \& {Chael}, A.~A. 2019{\natexlab{a}}, \apj, 881, 62

\bibitem[{{Palumbo} {et~al.}(2019{\natexlab{b}}){Palumbo}, {Doeleman}, {Johnson}, {Bouman}, \& {Chael}}]{Palumbo_2019}
{Palumbo}, D. C.~M., {Doeleman}, S.~S., {Johnson}, M.~D., {Bouman}, K.~L., \& {Chael}, A.~A. 2019{\natexlab{b}}, \apj, 881, 62

\bibitem[{{Palumbo} \& {Wong}(2022)}]{Palumbo:2022}
{Palumbo}, D. C.~M. \& {Wong}, G.~N. 2022, \apj, 929, 49

\bibitem[{{Palumbo} {et~al.}(2023){Palumbo}, {Wong}, {Chael}, \& {Johnson}}]{Palumbo:2023}
{Palumbo}, D. C.~M., {Wong}, G.~N., {Chael}, A., \& {Johnson}, M.~D. 2023, \apjl, 952, L31

\bibitem[{{Prussing}(1992)}]{Prussing:1992}
{Prussing}, J.~E. 1992, Journal of Guidance Control Dynamics, 15, 1037

\bibitem[{{Psaltis} {et~al.}(2018){Psaltis}, {Johnson}, {Narayan}, {Medeiros}, {Blackburn}, \& {Bower}}]{Psaltis:2018}
{Psaltis}, D., {Johnson}, M., {Narayan}, R., {et~al.} 2018, arXiv e-prints, arXiv:1805.01242

\bibitem[{{Psaltis} {et~al.}(2015){Psaltis}, {{\"O}zel}, {Chan}, \& {Marrone}}]{Psaltis:2015}
{Psaltis}, D., {{\"O}zel}, F., {Chan}, C.-K., \& {Marrone}, D.~P. 2015, \apj, 814, 115

\bibitem[{Pushkarev {et~al.}(2013)}]{Pushkarev:2013}
Pushkarev, A.~B. {et~al.} 2013, Astron. Astrophys., 555, A80

\bibitem[{Raymond {et~al.}(2024)Raymond, Doeleman, Asada, Blackburn, Bower, Bremer, Broguiere, Chen, Crew, Dornbusch, {et~al.}}]{Raymond:2024}
Raymond, A.~W., Doeleman, S.~S., Asada, K., {et~al.} 2024, The Astronomical Journal, 168, 130

\bibitem[{{Ricarte} \& {Dexter}(2015)}]{Ricarte:2015}
{Ricarte}, A. \& {Dexter}, J. 2015, \mnras, 446, 1973

\bibitem[{{Roelofs} {et~al.}(2019){Roelofs}, {Falcke}, {Brinkerink}, {Mo{\'s}cibrodzka}, {Gurvits}, {Martin-Neira}, {Kudriashov}, {Klein-Wolt}, {Tilanus}, {Kramer}, \& {Rezzolla}}]{Roelofs:2019}
{Roelofs}, F., {Falcke}, H., {Brinkerink}, C., {et~al.} 2019, \aap, 625, A124

\bibitem[{{Shavelle} \& {Palumbo}(2024)}]{Shavelle:2024}
{Shavelle}, K.~M. \& {Palumbo}, D. C.~M. 2024, \apjl, 970, L24

\bibitem[{{Shlentsova} {et~al.}(2024){Shlentsova}, {Roelofs}, {Issaoun}, {Davelaar}, \& {Falcke}}]{Shlentsova:2024}
{Shlentsova}, A., {Roelofs}, F., {Issaoun}, S., {Davelaar}, J., \& {Falcke}, H. 2024, \aap, 686, A154

\bibitem[{Smirnova {et~al.}(2014)Smirnova, Shishov, Popov, Gwinn, Anderson, Andrianov, Bartel, Deller, Johnson, Joshi, Kardashev, Karuppusamy, Kovalev, Kramer, Soglasnov, Zensus, \& Zhuravlev}]{Smirnova:2014}
Smirnova, T.~V., Shishov, V.~I., Popov, M.~V., {et~al.} 2014, \apj, 786, 115

\bibitem[{Smith(2024)}]{nasa_smex}
Smith, D. 2024, Astrophysics and Heliophysics Explorers Program

\bibitem[{{Tamar} {et~al.}(2024){Tamar}, {Hudson}, \& {Palumbo}}]{Tamar:2024A}
{Tamar}, A., {Hudson}, B., \& {Palumbo}, D. C.~M. 2024, \apj, 967, 90

\bibitem[{{Tamar} \& {Palumbo}(2024)}]{Tamar:2024}
{Tamar}, A. \& {Palumbo}, D. C.~M. 2024, \apj, 977, 147

\bibitem[{{The Event Horizon Telescope Collaboration}(2024)}]{EHT:2024_mid}
{The Event Horizon Telescope Collaboration}. 2024, arXiv e-prints, arXiv:2410.02986

\bibitem[{{Thompson} {et~al.}(2017){Thompson}, {Moran}, \& {Swenson Jr}}]{Thompson}
{Thompson}, A.~R., {Moran}, J.~M., \& {Swenson Jr}, G.~W. 2017, {Interferometry and Synthesis in Radio Astronomy}, 3rd edn. (Springer Open)

\bibitem[{{Tiede} {et~al.}(2022){Tiede}, {Johnson}, {Pesce}, {Palumbo}, {Chang}, \& {Galison}}]{Tiede:2022}
{Tiede}, P., {Johnson}, M.~D., {Pesce}, D.~W., {et~al.} 2022, Galaxies, 10, 111

\bibitem[{{Trippe} {et~al.}(2023){Trippe}, {Jung}, {Lee}, {Wagner}, {Han}, {Kang}, {Kyeong}, {Oh}, {Kim}, {Park}, \& {Hodgson}}]{Trippe:2023}
{Trippe}, S., {Jung}, T., {Lee}, J.-W., {et~al.} 2023, arXiv e-prints, arXiv:2304.06482

\bibitem[{{Tuntsov} {et~al.}(2013){Tuntsov}, {Bignall}, \& {Walker}}]{Tuntsov:2013}
{Tuntsov}, A.~V., {Bignall}, H.~E., \& {Walker}, M.~A. 2013, \mnras, 429, 2562

\bibitem[{{Vallado}(2013)}]{Vallado}
{Vallado}, D.~A. 2013, {Fundamentals of Astrodynamics}, 4th edn. (Microcosm Press)

\bibitem[{{Wang} {et~al.}(2024){Wang}, {Bilyeu}, {Boroson}, {Caplan}, {Riesing}, {Robinson}, {Schieler}, {Johnson}, {Blackburn}, {Haworth}, {Houston}, {Issaoun}, {Palumbo}, {Richards}, {Srinivasan}, {Weintroub}, \& {Marrone}}]{Wang:2024}
{Wang}, J., {Bilyeu}, B., {Boroson}, D., {et~al.} 2024, arXiv e-prints, arXiv:2406.09572

\bibitem[{{Weisskopf} {et~al.}(2022){Weisskopf}, {Soffitta}, {Baldini}, {Ramsey}, {O'Dell}, {Romani}, {Matt}, {Deininger}, {Baumgartner}, {Bellazzini}, {Costa}, {Kolodziejczak}, {Latronico}, {Marshall}, {Muleri}, {Bongiorno}, {Tennant}, {Bucciantini}, {Dovciak}, {Marin}, {Marscher}, {Poutanen}, {Slane}, {Turolla}, {Kalinowski}, {Di Marco}, {Fabiani}, {Minuti}, {La Monaca}, {Pinchera}, {Rankin}, {Sgro'}, {Trois}, {Xie}, {Alexander}, {Allen}, {Amici}, {Andersen}, {Antonelli}, {Antoniak}, {Attin{\`a}}, {Barbanera}, {Bachetti}, {Baggett}, {Bladt}, {Brez}, {Bonino}, {Boree}, {Borotto}, {Breeding}, {Brienza}, {Bygott}, {Caporale}, {Cardelli}, {Carpentiero}, {Castellano}, {Castronuovo}, {Cavalli}, {Cavazzuti}, {Ceccanti}, {Centrone}, {Citraro}, {D'Amico}, {D'Alba}, {Di Gesu}, {Del Monte}, {Dietz}, {Di Lalla}, {Persio}, {Dolan}, {Donnarumma}, {Evangelista}, {Ferrant}, {Ferrazzoli}, {Ferrie}, {Footdale}, {Forsyth}, {Foster}, {Garelick}, {Gunji}, {Gurnee}, {Head}, {Hibbard}, {Johnson}, {Kelly}, {Kilaru}, {Lefevre},
  {Roy}, {Loffredo}, {Lorenzi}, {Lucchesi}, {Maddox}, {Magazzu}, {Maldera}, {Manfreda}, {Mangraviti}, {Marengo}, {Marrocchesi}, {Massaro}, {Mauger}, {McCracken}, {McEachen}, {Mize}, {Mereu}, {Mitchell}, {Mitsuishi}, {Morbidini}, {Mosti}, {Nasimi}, {Negri}, {Negro}, {Nguyen}, {Nitschke}, {Nuti}, {Onizuka}, {Oppedisano}, {Orsini}, {Osborne}, {Pacheco}, {Paggi}, {Painter}, {Pavelitz}, {Pentz}, {Piazzolla}, {Perri}, {Pesce-Rollins}, {Peterson}, {Pilia}, {Profeti}, {Puccetti}, {Ranganathan}, {Ratheesh}, {Reedy}, {Root}, {Rubini}, {Ruswick}, {Sanchez}, {Sarra}, {Santoli}, {Scalise}, {Sciortino}, {Schroeder}, {Seek}, {Sosdian}, {Spandre}, {Speegle}, {Tamagawa}, {Tardiola}, {Tobia}, {Thomas}, {Valerie}, {Vimercati}, {Walden}, {Weddendorf}, {Wedmore}, {Welch}, {Zanetti}, \& {Zanetti}}]{Weisskopf:2022}
{Weisskopf}, M.~C., {Soffitta}, P., {Baldini}, L., {et~al.} 2022, Journal of Astronomical Telescopes, Instruments, and Systems, 8, 026002

\bibitem[{{Winkler} {et~al.}(2003){Winkler}, {Courvoisier}, {Di Cocco}, {Gehrels}, {Gim{\'e}nez}, {Grebenev}, {Hermsen}, {Mas-Hesse}, {Lebrun}, {Lund}, {Palumbo}, {Paul}, {Roques}, {Schnopper}, {Sch{\"o}nfelder}, {Sunyaev}, {Teegarden}, {Ubertini}, {Vedrenne}, \& {Dean}}]{Winkler:2003}
{Winkler}, C., {Courvoisier}, T.~J.~L., {Di Cocco}, G., {et~al.} 2003, \aap, 411, L1

\bibitem[{Xu \& Zhang(2020)}]{Xu:2020}
Xu, S. \& Zhang, B. 2020, Astrophys. J. Lett., 898, L48

\bibitem[{{Yuan} {et~al.}(2022){Yuan}, {Wang}, \& {Yang}}]{Yuan:2022}
{Yuan}, F., {Wang}, H., \& {Yang}, H. 2022, \apj, 924, 124

\bibitem[{{Zhu} {et~al.}(2019){Zhu}, {Johnson}, \& {Narayan}}]{Zhu:2019}
{Zhu}, Z., {Johnson}, M.~D., \& {Narayan}, R. 2019, \apj, 870, 6

\end{thebibliography}

\begin{appendix}

\section{Terminology}
The contents of this paper lie at the intersection of space mission design and astrophysical considerations pertinent to VLBI observations. Therefore, to improve accessibility, a simplified description of the key terms is given in Table \ref{t:terminology}.
\begin{table*}\label{t:terminology}
\caption{A glossary of the key terms and mathematical quantities used in the paper.}
    \centering
    \begin{tabular}{|c|c|}
    \hline
        Term & Description \\
        \hline
        Argument of Perigee $(\omega)$ & Orientation of the ellipse in the orbital
plane measured from the ascending node
to the periapsis.\\
Baseline & Vector drawn between two telescopes observing 
the same source orthographically
projected to the source. \\
        Eccentricity $(e)$ & Elliptical shape of the orbit $(0 \leq e < 1)$. \\
        Delta-v ($\Delta v)$ & The change in velocity required to perform a given manoeuvre.\\
         Diffractive Taper & A model for Sgr A*'s diffractive scattering effects.\\
         Downlink & A link from orbiter to an Earth station to perform data transfer.\\
         Inclination $(i)$ & Orientation of the orbit with respect to the
equator.\\
         Semi-major axis $(a)$ & Size of the orbit (average of the apoapsis
and periapsis radii).\\
         Specific Impulse $(I_{sp})$ & A measure of thrust accumulated by a rocket as fuel is burnt.\\
         True Anomaly $(\nu)$ & Denotes the angular position of the satellite with respect to the peripasis.\\
         \hline
    \end{tabular}

\end{table*}

\section{Ground array stations}
The Earth Centered Inertial (ECI) coordinates $(X,Y,Z)$ of the ground stations as part of the putative EHT 2025 array that are used to generate the $(u,v)$ coverage are given in Table \ref{t:ground+dlink}.
\begin{table}[h!] \label{t:ground+dlink}
\caption{The coordinates in an ECI frame of the VLBI ground array and the Downlink stations used to obtain the $(u,v)$ coverage.}
    \centering
    \begin{tabular}{|c|c|c|c|}
    \hline
    Station & X(km) & Y(km) & Z(km)\\
    \hline\hline
    \multicolumn{4}{|c|}{VLBI Ground Array (EHT 2025)}\\ 
    \hline
     ALMA & 2225.061 & -5440.057 & -2481.681\\
     APEX & 2225.039 & -5441.197   & -2479.303\\
     CARMA & -2397.431 & -4482.018 &   3843.524\\
     GLT   & 1500.692   &  -1191.735 &    6066.409\\
     JCMT  &-5464.584 & -2493.001 &   2150.653 \\
     KP & -1995.678 &  -5037.317 &  3357.328\\
     LMT   & -768.713 & -5988.541 &  2063.275 \\
     PDB & 4523.998   & 468.045 &    4460.309\\
     PV  & 5088.967 & -301.681 &   3825.015 \\
     SMA  & -5464.523 & -2493.147 &   2150.611 \\
     SMT & -1828.796 & -5054.406 &   3427.865 \\
     SPT & 0.0     &     0.0    &     -6359.609\\
     \hline
     \multicolumn{4}{|c|}{Downlink Ground Stations}\\ 
    \hline
    Cerro Paranal & 1946.434 & -5467.640 & -2642.704\\
    Haleakala & -5466.003 & -2404.290 & 2242.294\\
    Nemean & 4654.281 & 1947.909 & 3888.707\\
    Perth & -2384.691 & 4860.073 & -3361.166\\
    \hline
    \end{tabular}

\end{table}

\section{Orbit transfers and fuel budget considerations} \label{sec:Transfer_Fuel}
The paper considers two propulsion choices, namely CP and EP. As mentioned in the main text, in order for our results to be of relevance to technologies already being used in missions by NASA, we consider a cryogenic liquid methane and liquid oxygen (LOX) propellant for studying the Hohmann transfer via CP and the 4.5 kW SPT-140 Hall effect thruster which uses Xenon as a propellant, for an electric orbit raising (EOR) scheme. The CP transfer is envisaged to be performed using Falcon 9 Block 5 vehicle, which is the vehicle widely used in NASA's SMEX class missions. 

\subsection{Chemical propulsion}
Now, for the CP propulsion specifications, there are two main parameters of interest: the specific impulse $I_{sp}$ (usually measured in seconds (s)) which measures the efficiency of an engine to convert propellant to thrust, and $\Delta v$ (measured in km/s), which is the change in velocity required to perform the maneuver. For specified propellant, the $I_{sp}$ is 348s. Each of the orbit transfers are modelled as a Hohmann transfer and the $\Delta v$ for each of them is obtained using the \texttt{Maneuver} class of the Python package \texttt{poliastro}\footnote{\url{https://docs.poliastro.space/en/stable/}}. Rounded up to three decimal places, these are: 
\begin{gather}
    \Delta v_{7\rightarrow 13}= 1.962\, \text{km/s}, \nonumber\\ 
    \Delta v_{13\rightarrow 19}= 0.949\, \text{km/s}, \nonumber \\
    \Delta v_{19\rightarrow BX} = 0.702\, \text{km/s} \label{eq:delta_vs},
\end{gather}
wherein the subscripts denote the semi-major axes (in powers of $10^{3}$ km) between which the transfer is taking place, with $BX$ being $a=26563.88$km for the BHEX mission.
The total $\Delta v$ for the Hohmann transfer, denoted by $\Delta v_{t,H}$, is then given by, 
\begin{gather}
    \Delta v_{t,H}=v_{7\rightarrow 13}+v_{13\rightarrow 19}+v_{19\rightarrow BX}=3.613\, \text{km/s}. \label{eq:tot_deltav}
\end{gather}
Then, if $m_{0}$ is the initial mass of the payload with the propellant (``wet mass'') and $m_{f}$ is the final mass without the propellant (``dry mass''), then using the parameters given above, the latter can be computed from the former using the ideal rocket equation \citep{Vallado},
\begin{gather}
    m_{f}=\frac{m_{0}}{e^{\Delta v/(I_{sp}g)}}, \label{eq:rocket_eqn}
\end{gather}
where $g=9.81m/s^{2}$ is the standard gravitational acceleration. Thereafter, total propellant fuel required to perform the transfer is given by,
\begin{gather}
m_{\text{tot},\text{H}}=m_{0}-m_{f}. \label{eq:prop_reqd}
\end{gather}
For this paper, we consider $m_{0}=300$kg, which is around the maximum payload permissible in NASA's SMEX class missions. Finally, using this dry mass in the rocket Equation \ref{eq:rocket_eqn} along with the specified propellant $I_{sp}$ value, we can compute $m_{f}$ for the total $\Delta{v}$ given in \ref{eq:tot_deltav}. The propellant required for each maneuver is then given by Equation \ref{eq:prop_reqd}. Performing this analysis gives us the total propellant $m_{\text{tot},\text{H}}$ required for going from the Parking to the Target orbit using three Hohmann transfers: \begin{equation}\label{eq:total_prop_CP}
\begin{split}
m_{\text{tot},\text{H}}&=195.889\,\text{kg}.
\end{split}
\end{equation}
For subsequent comparison with the EP case, we also note the total time $t_{\text{tot},H}$ required in performing the three Hohmann transfers, obtained once again using the \texttt{Maneuver} class of \texttt{poliastro}:
\begin{equation}    
\begin{split}
    t_{\text{tot},H}&=t_{7\rightarrow13}+t_{13\rightarrow 19}+t_{19\rightarrow BX},\nonumber \\&=(6433.136+4802.079+21934.737)(\text{s}),\\&=9.21\, \text{hours}. \label{eq:t_hohmann} 
\end{split}
\end{equation}
\subsection{Electric propulsion}
\begin{figure} 
\centering
\includegraphics[width=\columnwidth]{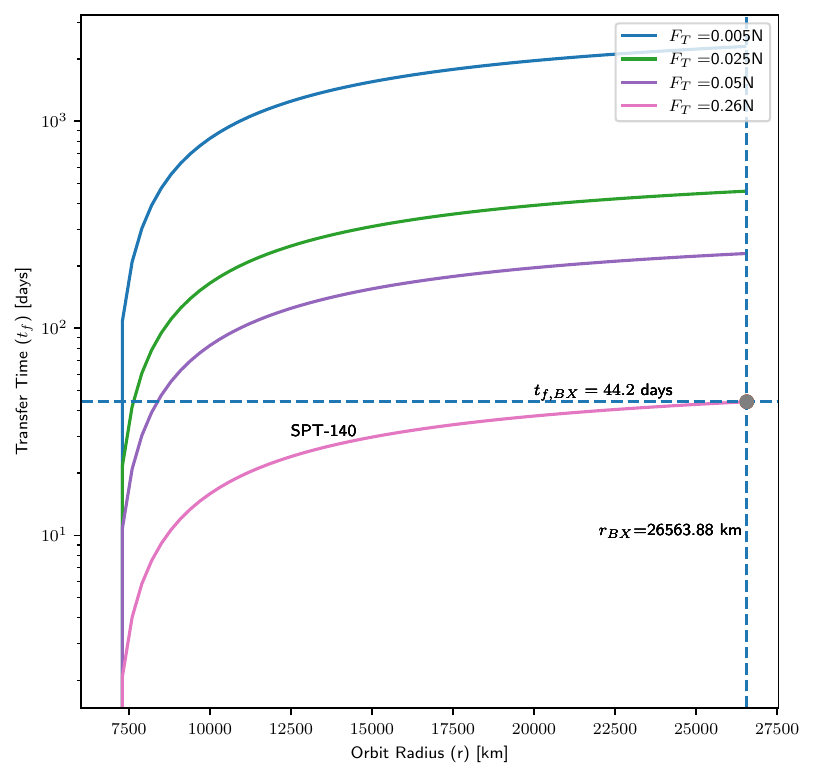}
\caption{The total transfer time taken for an EP-based orbit raising for various values of specific thrust $F_{T}$, as a function of orbital radius. A fixed value of $I_{sp}=1770$s is chosen which, along with $F_{T}=0.26 N$ are the specifications of the SPT-140 thruster. The time $t_{t,BX}$ is the time required to perform the orbit migration scheme of Parking$\rightarrow$IC1$\rightarrow$IC2$\rightarrow$Target developed in this paper.}
\label{f:EP_time}
\end{figure}

Since EP is not an impulsive burn rather a continuous thrust operation, the rocket Equation \ref{eq:rocket_eqn} does not hold exactly. Now, for a given electric propulsion thrust $F$ and $I_{sp}$, we first define the initial acceleration $a_{\mathcal{T},I}$ and the specific mass flow rate, 
\begin{gather}
    a_{T,I}=\frac{F}{m_{0}},\\
    \dot{m}=-\frac{-F/(g*I_{sp})}{m_{0}}. \label{eq:mdot}
\end{gather}
Then, for an initial orbit with radius $r_{\rm0}$ and period $T_{0}$, and final orbit with radius $r_{\rm f}$, the accumulated $\Delta v$ in terms of the ratio $R=r_{1}/r_{\rm0}$ is given by,
\begin{gather}
    \Delta v_{acc}=\Bigg(1-\sqrt{\frac{1}{R}}\Bigg)\Big(r_{0}{T_{0}}\Big).
\end{gather}
Finally, the total transfer time is given by Equation 6-44 of \cite{Vallado},
\begin{gather}
    t_{f}=\frac{1}{-\dot{m}}\Bigg(1-\exp\bigg(\frac{\dot{m}\Delta v_{acc}}{a_{T,I}}\bigg)\Bigg) \label{eq:time_ep}.
\end{gather}
This total time can then be used to compute the total mass $m_{f}$ of propellant required through the equation, 
\begin{gather}
    m_{f}=-\dot{m}t_{f}. \label{eq:prop_mass}
\end{gather}
Now, using Equation \ref{eq:time_ep}, the time taken to transfer from $r_{0}=7000$km to $r_{1}=26563.88$km assuming the SPT-140 parameters of $I_{sp}=1770$s and $F_{T}=0.26$N, is:
\begin{gather}
t_{f,BX}=44.2\, \text{days}.
    \label{eq:trans_time_EP}
\end{gather}
Then, the total propellant mass required, using Equation \ref{eq:prop_mass}, is:
\begin{gather}
m_{\text{tot},EP}=57.20\,\text{kg}\label{eq:total_prop_EP}.
\end{gather}
From the propellant requirements obtained in Equations \ref{eq:total_prop_CP} and  \ref{eq:total_prop_EP}, it is patently evident that EP leads to significant improvement in the overall fuel budget of the mission. However, as can be noted from comparing the time required for performing the CP-based Hohmann transfer and the EP-based EoR, the savings in propellant mass comes at a cost of relatively longer transfer time. Lastly, Figure \ref{f:EP_time} plots the total time taken to achieve a given orbital radius for varying values of the specific thrust $F_{T}$. It is evident that as $F_{T}$ increases, the transfer time reduces. Since SPT-140 has a relatively higher $F_{T}$, it has a reasonable transfer time of 44.2 days to reach the proposed BHEX target orbit with radius $r_{\rm BX}=26563.88$ km.

The specific impulse parameter ($I_{sp}$) deserves some clarification since its units of seconds might seem a little confusing. Simply, specific impulse measures the efficiency of an engine by computing how much thrust it can produce per unit of the propellant. In particular, the phrase ``specific'' here means divided by the weight $(w)$ (of the fuel) and the quantity being divided is the impulse $I$ of the thruster, with the usual units of Newtons-second. This impulse is defined in terms of the mass of the fuel being expelled and the effective exhaust velocity of the thruster \citep{Vallado} but for the current discussions, these specifications are not necessary; we just need to recall the SI units of impulse. Thus, using standard dimensional analysis, we can obtain the dimensions of $I_{sp}$ as:
\begin{gather}
    [I_{sp}]=\frac{[I]}{[w]}=\frac{MLT^{-1}}{MLT^{-2}}=T
\end{gather}
with $T$ being the time dimension which in SI units is measured in seconds.

\end{appendix}

\end{document}